\documentclass[sigconf]{acmart}
\usepackage{amsmath,amsfonts}
\usepackage{graphicx}
\usepackage{textcomp}
\usepackage{xspace}
\usepackage{algorithm}
\usepackage{algpseudocode}
\usepackage{fancyhdr}
\usepackage[normalem]{ulem}
\usepackage{booktabs}
\usepackage{multirow}
\usepackage{setspace}
\usepackage{siunitx}
\usepackage{titlesec}
\usepackage{tikz}
\usepackage{enumitem}
\usepackage{pifont}
\usepackage{mathtools}
\usepackage{dblfloatfix} 
\usepackage[bottom]{footmisc}
\usepackage{balance}
\usepackage{xargs}  
\usepackage{setspace}
\usepackage{datetime}
\usepackage{float}

\newcommand{\namePaper}{{Venice}\xspace}
\newcommand{\pkt}{{scout packet}\xspace}
\newcommand{\pkts}{{scout packets}\xspace}

\newcommand{\ideal}{path-conflict-free SSD}

\newcommand{\baseline}{Baseline SSD}
\newcommand{\Baseline}{Baseline SSD}
\newcommand{\nossd}{NoSSD}
\newcommand{\pssd}{pSSD}
\newcommand{\pnssd}{pnSSD}
\newcommand{\conf}{path conflict}
\newcommand{\confs}{path conflicts}

\newcommand{\Confs}{Path conflicts}

\newcommand{\tbers}{\texttt{tBERS}\xspace}
\newcommand{\tr}{\texttt{tR}\xspace}
\newcommand{\tprog}{\texttt{tPROG}\xspace}

\definecolor{darkpastelgreen}{rgb}{0.01, 0.75, 0.24}
\definecolor{darkspringgreen}{rgb}{0.09, 0.45, 0.27}
\definecolor{applegreen}{rgb}{0.55, 0.71, 0.0}

\newcommand{\dq}[1]{{\color{black}#1}}
\newcommand{\reva}[1]{{\color{black}#1}}
\newcommand{\revb}[1]{{\color{black}#1}}

\newcommand{\tab}[1]{{Table~#1}} %
\newcommand{\head}[1]{{\noindent\textbf{#1.}}} %

\newcommand{\nika}[1]{{\color{black}#1}}

\usepackage[colorinlistoftodos,prependcaption,textsize=scriptsize]{todonotes} %
\usepackage{marginnote} 
\let\marginpar\marginnote
\newcommand{\circled}[1]{{\tikz[baseline=(char.base)]{\node[shape=circle,inner sep=1.3pt,fill=black, text=white, scale=0.8] (char) {\small \textbf{#1}};}}}
\newcommand{\whitecircled}[1]{{\tikz[baseline=(char.base)]{\node[shape=circle, draw, inner sep=1.3pt,fill=white, text=black, scale=0.8] (char) {\small \textbf{#1}};}}}

\newcommand{\orevII}[1]{{\color{black}#1}}
\newcommand{\orevIII}[1]{{\color{black}#1}}
\newcommand{\orevIV}[1]{{\color{black}#1}}
\newcommand{\orevV}[1]{{\color{black}#1}}
\newcommand{\orevVI}[1]{{\color{black}#1}}
\newcommand{\orevVII}[1]{{\color{black}#1}}

\newcommand{\vetII}[1]{{\color{black}#1}}
\newcommand{\vetIV}[1]{{\color{black}#1}}

\newcommand{\algcomment}[1]{\textcolor{blue}{//#1}}

\AtBeginDocument{%
  \providecommand\BibTeX{{%
    \normalfont B\kern-0.5em{\scshape i\kern-0.25em b}\kern-0.8em\TeX}}}

\copyrightyear{2023} 
\acmYear{2023} 
\setcopyright{acmlicensed}
\acmConference[ISCA '23]{Proceedings of the 50th Annual International Symposium on Computer Architecture}{June 17--21, 2023}{Orlando, FL, USA}
\acmBooktitle{Proceedings of the 50th Annual International Symposium on Computer Architecture (ISCA '23), June 17--21, 2023, Orlando, FL, USA}
\acmPrice{15.00}
\acmDOI{10.1145/3579371.3589071}
\acmISBN{979-8-4007-0095-8/23/06}

\sloppypar

\begin{document}

\title{Venice: Improving Solid-State Drive Parallelism \\ at Low Cost via Conflict-Free Accesses}

\newcommand{\affilETH}[0]{\textsuperscript{\S}}
\newcommand{\affilSUT}[0]{\textsuperscript{$\dagger$}}
\newcommand{\affilIPM}[0]{\textsuperscript{$\ddagger$}}
\newcommand{\affilPostec}[0]{\textsuperscript{$\nabla$}}

\author{*Rakesh Nadig\affilETH \hspace{0.5cm} *Mohammad Sadrosadati\affilETH \hspace{0.5cm}  Haiyu Mao\affilETH \hspace{0.5cm}  
Nika Mansouri Ghiasi\affilETH \hspace{0.5cm}  \\ Arash Tavakkol\affilETH \hspace{0.5cm} Jisung Park\affilETH\affilPostec \hspace{0.5cm}  Hamid Sarbazi-Azad\affilSUT\affilIPM \hspace{0.5cm}    Juan G{\'o}mez Luna\affilETH \hspace{0.5cm} Onur Mutlu\affilETH \\ \vspace{0.2cm}
\emph{\affilETH ETH Z{\"u}rich \hspace{1cm} \affilPostec POSTECH \hspace{1cm}  \affilSUT Sharif University of Technology \hspace{1cm} \affilIPM IPM \vspace{0.1cm}\\}
\thanks{*Rakesh Nadig and Mohammad Sadrosadati are co-primary authors.}
}

\renewcommand{\shortauthors}{R. Nadig and M. Sadrosadati, et al.}

\renewcommand{\authors}{Rakesh Nadig, Mohammad Sadrosadati, Haiyu Mao, Nika Mansouri Ghiasi, Arash Tavakkol, Jisung Park, Hamid Sarbazi-Azad, Juan G{\'o}mez-Luna, Onur Mutlu}

\begin{abstract}
The performance and capacity of solid-state drives (SSDs) are continuously improving to meet the increasing demands of modern data-intensive applications. Unfortunately, communication between the SSD controller and memory  chips (e.g., 2D/3D NAND flash chips) is a critical performance bottleneck for many applications. SSDs use a multi-channel shared bus architecture where multiple memory chips connected to the same channel communicate to the SSD controller with only one path. As a result, \confs~\orevII{often} occur during the servicing of multiple I/O requests, \orevII{which} significantly limits SSD parallelism.  
It is critical to handle path conflicts well to improve SSD parallelism and performance. 

Our goal is to fundamentally tackle the \conf~problem by increasing the number of paths between the SSD controller and memory chips at low cost. \orevII{To this end, we build on the idea of using an interconnection network to increase the path diversity between the SSD controller and memory chips}. We propose \emph{\namePaper{}}, a new mechanism that introduces a low-cost interconnection network between the SSD controller and memory chips and utilizes the path diversity to intelligently resolve path conflicts. \namePaper{} employs three key techniques: 1) a simple router chip added next to each memory chip \emph{without} modifying the memory chip design, 2) a path reservation technique that reserves a path from the SSD controller to the target memory chip before initiating \orevII{a} transfer, %
and 3) a fully-adaptive routing algorithm \orevII{that effectively utilizes} the path diversity \orevII{to} resolve path conflicts. Our experimental results show that \namePaper{} 1) improves performance by an average of 2.65$\times$/1.67$\times$ over a baseline performance-optimized/cost-optimized SSD design across a wide range of workloads, 2) \orevII{reduces energy consumption by an average of~\orevIII{61\%} compared to a baseline performance-optimized SSD design. Venice’s benefits come at a relatively
low area overhead.} %
\end {abstract}


\maketitle

\section{Introduction\label{sec:introduction}}
Flash-memory-based solid-state drives (SSDs) are ubiquitous, from cloud environments to mobile devices ~\cite{dirik-isca-2009, do2013query, chen2011hystor, stefanov2013oblivistore, micheloni2010inside,agrawal2008design, tavakkol2014design, eshghi2018ssd, hu2012exploring, cho2015design, gao2014exploiting, gao2017exploiting, kim2018autossd, tavakkol2016performance,kim2022networked,lim2010faster, kim2019practical,cho2013active,cai-errors-2018,
cai2015data,
cai2013threshold,
cai2017error,
tavakkol2018flin,
park2022flash,
mutlu-emergingcomputing-2021,
kim2020evanesco,
cai2017vulnerabilities,
cai2012error}. The high performance, low power consumption and shock resistance of SSDs make them suitable replacements for hard disk drives (HDDs)~\cite{dirik-isca-2009, kryder2009after, mielke2017reliability}. The rise in the number of data-intensive applications has resulted in the widespread adoption of SSDs in computing systems, increasing the demand for higher performance and capacity in SSDs. Although SSD vendors have significantly improved both performance and capacity of SSDs (e.g.,~\cite{znand,NVMExpress:2014,micron3dxpoint, shibata20191, cai2017error}) over the years, communication \orevII{\emph{within}} the SSD (i.e., between the SSD controller and NAND flash chips) is still a critical performance bottleneck \cite{grupp2012bleak,nishtala2015high,hsu2015nand,tavakkol2012network,tavakkol2014design,park2022flash,jung2012evaluation, jung2012physically,hu2011performance,park2010exploiting,mao2017improving,ruan2012improving,gao2019parallel,choi2018parallelizing} for many applications, especially workloads with a large number of random I/O requests \cite{min2012sfs, han2019wal, kim2022networked, tavakkol2014design, tavakkol2012network,zhou2015efficient, choi2018parallelizing,kim2019practical, agrawal2008design,kim2015improving}.

Commodity SSDs use a multi-channel shared bus architecture (e.g.,~\cite{agrawal2008design, tavakkol2014design, dirik-isca-2009, eshghi2018ssd, hu2012exploring, cho2015design, gao2014exploiting, gao2017exploiting, kim2018autossd, tavakkol2016performance,kim2022networked}) for communication between the SSD controller and NAND flash chips. In \orevII{this} architecture, the SSD controller is connected to flash chips via multiple channels (typically 4 to 16~\cite{samsung-ssd-980, skhynix-gold-p31, microchip-16-channel-controller}) with a number of flash chips (typically 4 to 32~\cite{samsung-ssd-980, skhynix-gold-p31, lee2022mqsim, pm9a3}) connected to each channel. Thus, each flash chip has only \emph{one path} to communicate with the SSD controller and several flash chips share the same path. 
As a result, there is a high likelihood that multiple I/O requests access NAND flash chips on the \emph{same} channel. \vetII{These I/O requests should be transferred serially on the same channel,} 
which significantly limits SSD parallelism.
We call this problem \emph{\conf}. To quantify the effect of \confs~on SSD performance, we compare the performance of a state-of-the-art baseline SSD with an \orevII{\emph{ideal
} \orevIII{(i.e., path-conflict-free)}} SSD. We observe that the ideal SSD outperforms the baseline SSD by \reva{4$\times$} on average across \reva{nineteen} data-intensive real-world workloads (see \S\ref{sec:motivation} for more detail).

SSD vendors attempt to reduce 
\confs~by increasing the number of channels in the SSD. However, this is not a scalable solution since increasing the number of channels makes
the SSD controller more complex (e.g., the SSD controller needs more I/O pins to service more parallel channels), 
increasing the \orevII{overall} cost of the SSD. A recent prior work~\cite{kim2022networked} attempts to address the \conf~problem by increasing the bandwidth of each SSD channel. \orevII{This work proposes} to utilize \orevII{the} control and data pins of flash chips for \orevII{both} command \orevII{and} data transfer, effectively providing  2$\times$ the SSD channel bandwidth. Unfortunately, such techniques \orevII{1)} are expensive as they require \orevII{relatively large} modifications to the \orevII{commodity} NAND flash \orevIII{memory} chips (e.g., 20\% area overhead \orevII{in} each flash die ~\cite{kim2022networked}), \orevII{and 2) alleviate but cannot \orevIII{effectively} resolve \confs} \orevIII{(as we show in \S\ref{subsec:motivation_effectiveness})}.

\textbf{Our goal} is to fundamentally address the \conf~problem in SSDs by providing high \emph{path diversity} at low cost for communication between the SSD controller and flash chips. Our key idea is to use a low-cost interconnection network to increase the path diversity between \orevII{the} SSD controller and flash chips. Some prior works propose the use of an %
interconnection network \orevII{within an} SSD to provide a scalable solution to increase SSD capacity~\cite{tavakkol2014design,tavakkol2012network}. Such works can potentially be repurposed to tackle the \conf~problem. 
However, prior SSD interconnection network designs have two main weaknesses, which prevent them from effectively addressing the \conf~problem.  First, prior works impose significant area \orevII{(i.e., cost)} overhead as they integrate a buffered router (e.g., 16KB buffer per router port) inside each flash chip. \orevII{Such a design} increases the area and the number of I/O pins of the flash chip. Second, prior works do \emph{not} resolve \confs~effectively \orevII{because they} employ a simple deterministic routing algorithm, which cannot utilize the interconnection network's rich path diversity \orevIII{(\orevIV{as we show in} \S\ref{subsec:motivation_effectiveness})}.

We propose \emph{\namePaper{},}\footnote{Named after the network of canals in \orevII{the city of} \namePaper{}~\cite{venice}.} a new mechanism that introduces a low-cost interconnection network of flash chips to fundamentally tackle the \conf~problem while effectively addressing the two \orevII{major weaknesses} of prior works on SSD interconnection networks. \namePaper{} employs three key techniques. First, \namePaper{} adds a new router chip \orevII{\emph{next to}} each flash chip \orevII{\emph{without}} modifying the flash chip. Routers are connected in a network topology, such as a 2D mesh. Second, 
\namePaper{} reserves a \orevII{network} path for each I/O request before initiating the \orevII{command and data} transfer. This technique ensures that the I/O request transfer does \orevII{\emph{not}} experience path conflicts in the network, which avoids the need for large buffers in each router. 
Third, to find a free path between the SSD controller and the flash chip, \namePaper{} uses a non-minimal fully-adaptive routing algorithm \orevII{that effectively utilizes} the interconnection network's path diversity. 

We evaluate \namePaper{} using MQSim~\cite{tavakkol2018mqsim,mqsim-github}, \orevII{a} state-of-the-art SSD simulator. We use two baseline SSD configurations, \emph{performance-optimized} and \emph{cost-optimized} \orevII{and a wide variety of I/O-intensive benchmarks (see \S\ref{sec:evaluation})}. \vetII{Our evaluation yields three key results that demonstrate \namePaper{}'s effectiveness.} 
First, for the performance-optimized configuration, %
\namePaper{} improves performance by an average of 2.65$\times$ (\vetII{up to 7.10$\times$}) and 1.92$\times$ (\vetII{up to 4.30$\times$}) compared to the baseline SSD design and best-performing prior work~\cite{tavakkol2012network} (without taking the overhead of prior work into account), respectively. For the cost-optimized configuration, \namePaper{} improves performance by an average of 1.67$\times$ \vetII{(up to 3.68$\times$)} and 1.47$\times$ \vetII{(up to 2.90$\times$)} compared to the baseline SSD design and the best-performing prior work~\cite{tavakkol2012network}, respectively. \vetII{Second,} 
\orevII{\namePaper{} reduces energy consumption by \vetII{an average of} 61\% and \vetII{46\%} compared to the baseline performance-optimized SSD design and the best-performing prior work~\cite{tavakkol2012network}, respectively.} %
\vetII{Third, \namePaper{}’s benefits come at a relatively low area overhead. \namePaper{}'s routers impose 8\% area overhead to the SSD  printed circuit board (PCB). \namePaper{}'s interconnection network links, in total, occupy 44\% \orevIII{lower} area compared to the baseline multi-channel \orevIII{shared} bus architecture.}

This paper makes the following contributions:
\begin{itemize}

\item \orevIII{We demonstrate the importance of the path conflict problem in modern SSD designs\orevIV{, and quantify its performance impact}.}

\item We propose \namePaper{}, a new mechanism that introduces a low-cost interconnection network of flash chips to fundamentally address the \conf~problem in SSDs.

\item We introduce three key techniques that enable \namePaper{}: 1) a simple router chip added next to each flash chip without modifying the flash chip itself, 2) a path reservation technique to reserve \orevII{paths} from the SSD controller to \orevII{target flash chips}, and 3) a non-minimal fully-adaptive routing algorithm to effectively utilize the path diversity in the interconnection network.

\item We rigorously evaluate \namePaper{} and show that it significantly improves performance over state-of-the-art \orevII{SSD} designs on both performance- and cost-optimized SSD configurations. 

\end{itemize}

\section{Background}
\label{sec:background}

\begin{figure}[h]
\centering
\includegraphics[width=\linewidth]{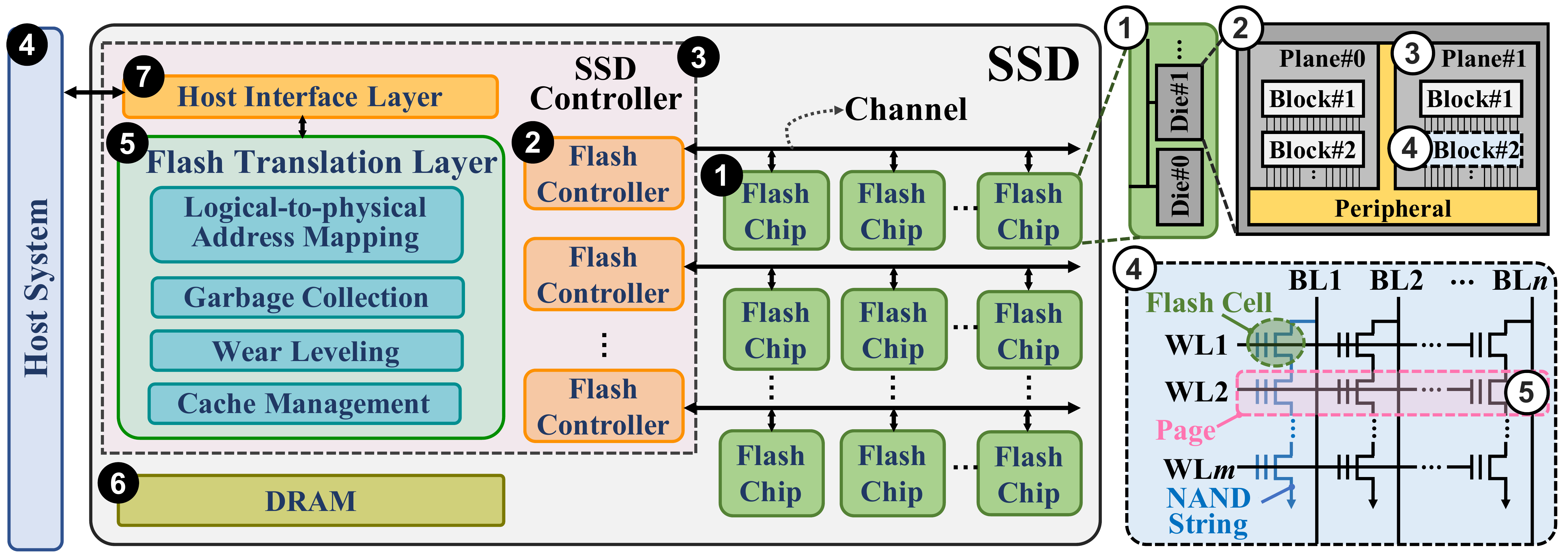}
\vspace{-6mm}
\caption{High-level overview of a modern SSD}
\label{fig:ssd}
\end{figure}

We provide a brief background on the baseline multi-channel shared bus SSD architecture. A typical modern SSD consists of an SSD controller and an array of flash chips. The host system uses a high-speed communication interface (e.g., PCIe~\cite{pcie-spec}) to communicate with the SSD. The SSD controller communicates with the flash chips using the shared flash channels. \orevII{Figure \ref{fig:ssd} shows a high-level overview of a modern SSD}. 

\subsection{Flash Chip Array}
Multiple flash chips (\circled{1})~\cite{cai2017error, agrawal2008design} are connected to a flash controller (\circled{2}) through a shared channel in the multi-channel shared bus architecture. Each flash chip (\whitecircled{1}) contains one or more \orevIV{(typically 1 to 4)} flash dies. 
\orevII{Flash dies operate independently of each other.} 
Each die (\whitecircled{2}) consists of multiple (e.g., 2 or 4) planes. 
A plane (\whitecircled{3}) typically contains thousands of blocks, and each block (\whitecircled{4}) consists of tens to hundreds of pages. 
\orevII{A flash page (\whitecircled{5}) \orevIV{consists of a set of flash cells connected to the same wordline within a flash block}. \orevIV{Read} and write operations are typically performed at the granularity of a flash page (e.g., 16KB in size). However, erase operations happen \orevIV{at} block granularity\orevIV{~\cite{cai-errors-2018, cai2015data,cai2013threshold,cai2017error,
park2022flash,cai2017vulnerabilities,cai2012error}}. Planes in the same die share the peripheral circuitry used to access pages; as such, they can
concurrently operate only when accessing pages (or blocks) at the
same offset, which are called \emph{multi-plane operations}.  
} 

Based on the number of bits stored in a flash cell, it is categorized as a single-level cell (SLC; 1 bit)~\cite{cho2001dual}, multi-level cell (MLC; 2 bits)~\cite{lee20043}, triple-level cell (TLC; 3 bits)~\cite{maejima2018512gb}, or quad-level cell (QLC; 4 bits)~\cite{cho20221}. 
The capacity of the SSD increases as the flash cell stores more bits, but the increased flash cell density leads to \orevII{higher latency} and \orevII{lower endurance}~\cite{grupp2012bleak, dirik-isca-2009, mohan2010learned,boboila2010write,cai2017vulnerabilities, cai2017error,micheloni2010inside, mielke2017reliability,cai2012error,jimenez2014wear,margaglia2015improving,cai-errors-2018}.
\vspace{-1mm}
\subsection{SSD Controller}
The SSD controller (\circled{3}) is responsible for managing the NAND flash chips (\circled{1}) and the I/O requests sent by the host (\circled{4}). 
\orevII{The} SSD controller contains an embedded microprocessor that executes firmware called \orevII{the} Flash Translation Layer (FTL) (\circled{5})~\cite{gupta2009dftl, tavakkol2018mqsim, lim2010faster, shin2009ftl, zhou2015efficient}. The SSD controller stores metadata (e.g., a logical-to-physical \orevII{page} mapping table) used to manage the FTL functionality and caches frequently accessed pages in DRAM (\circled{6}) that is part of the SSD. 
\orevII{The SSD controller consists of multiple flash controllers (\circled{2}). A flash controller is an embedded processor that interfaces with multiple flash chips using a shared channel. The flash controller selects the flash chip for a read/write operation and initiates the command and data transfer.}

\textbf{Host Interface Layer.}
Host Interface Layer (HIL) (\circled{7})~\cite{jung2019design, jung2014hios, tavakkol2018mqsim} is the interface between the host system (\circled{4}) and the SSD controller (\circled{3}). 
\orevII{HIL communicates with the host system using a communication protocol over the system I/O bus.}
\orevII{HIL in a commodity SSD typically supports the Advanced Host Controller Interface (AHCI)~\cite{ahci} or the NVM Express (NVMe)~\cite{NVMExpress:2014} interface. AHCI builds upon the Serial ATA (SATA)~\cite{sata-spec} protocol, which is commonly used to connect the host system to the hard disk drives. 
AHCI and SATA interfaces provide very low throughput for SSDs because of the availability of a single I/O queue to submit I/O requests to the SSD. 

To overcome the throughput bottleneck of AHCI and SATA, modern SSDs have adopted the NVMe protocol, which uses the PCI Express (PCIe) system bus to communicate with the host (see, e.g., ~\cite{IntelP45002017,TOSHIBA2016,HGSTSN2002017,SandDiskSkyhawk2017,OCZ2017,tavakkol2018flin,tavakkol2018mqsim, znand, pm9a3,980pro,990pro}).   
}
NVMe directly exposes multiple SSD I/O queues to the host, thereby enabling 1) high-bandwidth and low-latency communication between the SSD and the host, 2) more fine-grained control of the I/O request scheduling policy by the SSD controller~\cite{tavakkol2018flin}.
The host system transfers I/O requests into a \emph{Submission Queue} allocated to the application in the HIL.
HIL picks an I/O request from the \emph{Submission Queue} and sends the request to the FTL for processing. 
After the completion of the I/O request, the HIL updates the \emph{Completion Queue} to inform the host system. 

\textbf{Flash Translation Layer (\circled{3}).} 
FTL has four major responsibilities~\cite{gupta2009dftl, tavakkol2018mqsim, lim2010faster, shin2009ftl, zhou2015efficient}. 
First, for each page of data, FTL manages the mapping of \orevII{each} logical address (i.e., the \orevII{requested address in the} host system's address space) to \orevII{a} physical address (i.e., the actual location \orevII{in the physical flash chips} where the \orevII{requested} data resides).
\orevII{Before new data is written to a flash page, an entire flash block \vetII{that contains the target flash page} has to be erased. This is called the \emph{erase-before-write} requirement. Unfortunately, the \emph{erase-before-write}  requirement of NAND flash memory makes in-place writes \orevIV{prohibitively} costly in terms of performance, energy consumption, and lifetime
\cite{cai2017error, kim2020evanesco,cai2017vulnerabilities,gupta2009dftl,cai-errors-2018}. To overcome this issue, %
the FTL implements \orevII{an \emph{out-of-place}} write policy in modern SSDs %
~\cite{cai2017vulnerabilities, cai2017error,cai-errors-2018}.}
Whenever a page of data is written to by the host system \orevII{to a logical page address}, the FTL 1) invalidates the \orevII{corresponding} physical page address where the overwritten data resides, 2) writes the new \orevIV{page} data to a \orevII{\emph{different}} physical page address, %
and 3) %
updates the \orevII{logical-to-physical page mapping metadata} of \orevII{the logical} page.  

Second, the FTL performs \orevII{\emph{garbage collection (GC)}}~\cite{yang2014garbage, cai2017error, tavakkol2018mqsim, agrawal2008design, shahidi2016exploring, lee2013preemptible,jung2012taking, choi2018parallelizing,wu2016gcar,cai-errors-2018,cai2017vulnerabilities} to recover the wasted space due to pages invalidated by the out-of-place write policy. During GC, the FTL 1) chooses a victim block with the least number of valid pages, 2) copies all valid pages in the victim block to another block, 3) updates the logical to physical address mapping metadata for pages that have been already copied, and 4) erases the victim block to \orevII{use this block for future write operations}.
Third, the FTL implements a \emph{wear-leveling} technique to distribute the writes evenly across all the flash blocks so that the flash blocks in the SSD wear out in a uniform manner~\cite{shin2009ftl,lim2010faster,zhou2015efficient, murugan2011rejuvenator, agrawal2008design, jimenez2014wear}. 
Having a wear-leveling mechanism in FTL is critical for SSD lifetime as the number of times a flash block can be erased and programmed is limited~\cite{sun2016exploring, cai2017error,tan2015cost}. 
Fourth, the %
FTL avoids frequent lookups to the flash memory by caching frequently-accessed data (e.g., the logical-to-physical page mapping table~\cite{gupta2009dftl}) or \orevII{frequently-requested pages by the host} in the DRAM (\circled{6}) \orevIV{that is} present inside the SSD. %

\textbf{Flash \orevII{Controller} (\circled{2}).} 
A flash controller (FC)~\cite{kim2021decoupled, kim2022networked,wu2012reducing, micheloni2010inside, microchip-16-channel-controller} is an embedded processor in an SSD that interfaces with multiple flash chips connected through a shared channel.
The FTL communicates with the FC to perform a NAND flash operation. The FC communicates with the flash chips using the control/data and arbitration pins ~\cite{micheloni2010inside,onfi-2022}. 
For a write operation, the FC 1) performs data randomization to avoid high bit error rates caused by worst-case data patterns~\cite{cai2017error,cai2017vulnerabilities,cai-errors-2018}, 2) performs Error-Correcting Code (ECC) encoding to improve reliability and performance~\cite{cai2017error, cai2017vulnerabilities,park2022flash,zhao2013ldpc, micheloni2010inside,tanakamaru2013error,park2021reducing}, 3) sends \orevII{a} write command (\orevII{with the} physical page address), to the target flash chip, and 4) transfers the randomized ECC-encoded write data to the target flash chip. 
For a read operation, the FC 1) sends \orevII{a} read command to the target flash chip, 2) receives the read data from the flash chip, 3) performs ECC decoding and corrects possible errors in the data~\cite{cai2017error, cai2017vulnerabilities,park2022flash,zhao2013ldpc, micheloni2010inside,tanakamaru2013error,park2021reducing, cai-errors-2018},\footnote{The FC retries the read process if ECC decoding fails\orevII{~\cite{park2021reducing,cai-errors-2018,cai2017vulnerabilities,cai2017error,shim2019exploiting,cui2022improving,du2017laldpc,micheloni2010inside,liu2019soml,cai2015data,cai2012error,cai2013threshold}.}} and 4) derandomizes the read data to recover the original data. %

\section{Motivation\label{sec:motivation}}
 
We describe the \conf~problem in a typical multi-channel shared bus SSD architecture (which we call \orevII{\emph{\baseline{}}}) and \orevII{major} approaches to mitigate path conflicts. 

\subsection{The Path Conflict Problem in Modern SSDs\label{subsec:motivation_pathconflict}}
\begin{figure}[h]
\centering
\includegraphics[width=0.95\linewidth]{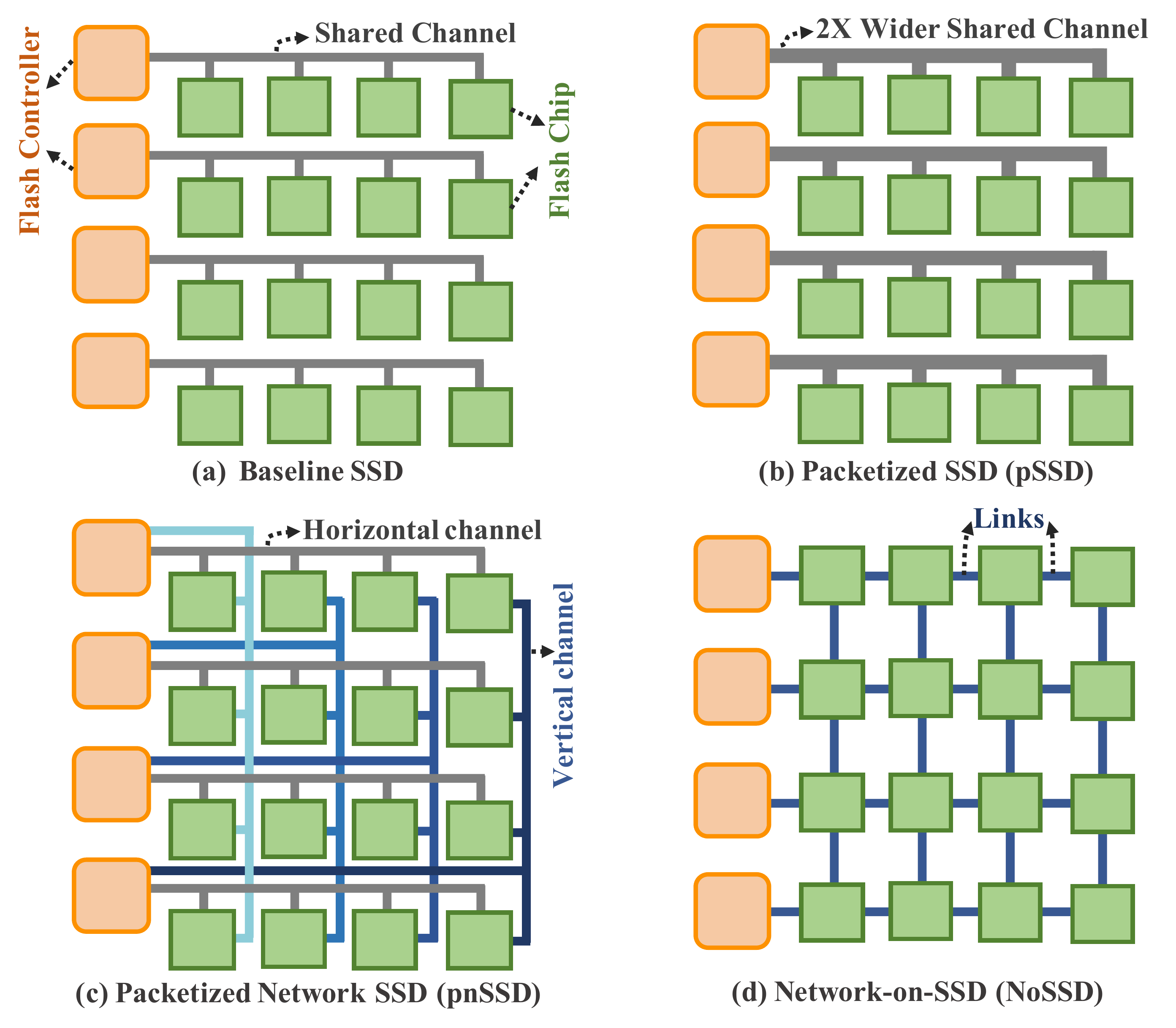}
\vspace{-3mm}
\caption{\orevII{Flash chip array architecture of four SSD designs: \baseline{}, Packetized SSD (\pssd{})~\cite{kim2022networked}, Packetized Network SSD (\pnssd{})~\cite{kim2022networked}, and Network-on-SSD (\nossd{})~\cite{tavakkol2012network}.}}
\label{fig:network}
\end{figure}

A typical SSD (e.g.,~\cite{agrawal2008design, dirik-isca-2009, gao2014exploiting,lee2022mqsim,pm9a3, 980pro, 990pro, eshghi2018ssd, hu2012exploring, cho2015design, kim2018autossd, tavakkol2016performance}) \orevII{uses a multi-channel shared bus architecture for communication between the SSD controller and NAND flash chips.} 
\orevII{The SSD controller is connected to flash chips via} a number of shared channels (\orevII{typically} 4 to 16~\cite{samsung-ssd-980, skhynix-gold-p31, microchip-16-channel-controller}) with multiple flash chips (\orevII{typically} 4 to 32~\cite{samsung-ssd-980, skhynix-gold-p31, lee2022mqsim, pm9a3}) connected to each channel. Figure~\ref{fig:network}(a) shows an example configuration in \orevII{such a} \baseline{} where there are four shared channels with four flash chips connected to each shared channel. In \baseline{}, each flash chip has \emph{only} one channel (or \emph{path}) to communicate with the SSD controller. Unfortunately, the flash chips connected to the same channel share the same path to the SSD controller. Path sharing \orevII{causes} the \conf~problem, where \orevII{an I/O request needs to wait for the path to become free, if the path is being used for another I/O request.}

\begin{figure}[h]
\centering
\includegraphics[width=\linewidth]{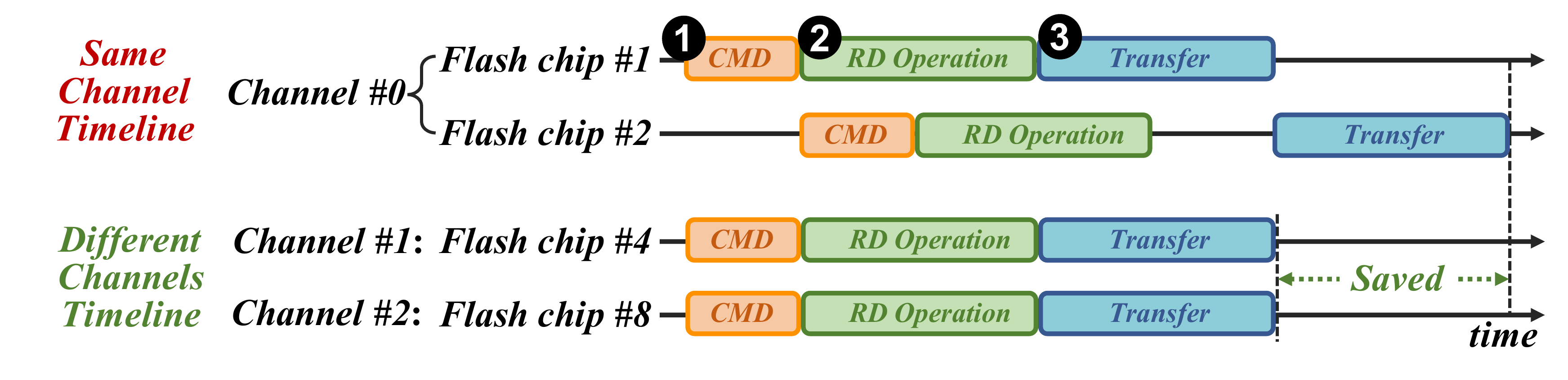}
\vspace{-6mm}
\caption{Service timeline of two read requests to two different flash chips. The \orevII{two} flash chips are connected to the same channel (top) or different channels (bottom).}
\label{fig:conflict}
\end{figure}

To \orevII{demonstrate} the \conf~problem, we show two examples of service timelines of ongoing read I/O requests in Figure~\ref{fig:conflict}. For simplicity, the figure shows only three major steps during a read request, the read command (\emph{CMD}~\circled{1}), the flash read operation (\emph{RD Operation}~\circled{2}), and read data transfer from \orevII{the} flash chip to the SSD controller (\emph{Transfer}~\circled{3}). 

The first example (\orevIV{Figure \ref{fig:conflict}} top) shows two ongoing read requests to two different flash chips connected to the \emph{same} channel (i.e., \orevII{the two requests experience the} \conf~problem). Unfortunately, in this case, only the second step (\circled{2}), flash read operation, of the two ongoing read requests can be performed in parallel. Other steps (\circled{1} and \circled{3}) should be performed one after the other \orevII{(i.e., serially)} because they use the same path, which increases the total service time (the total time taken for processing an I/O request within the SSD) of these two requests. 
The second example (\orevIV{Figure \ref{fig:conflict}} bottom) shows two ongoing read requests to two flash chips connected to two \emph{different} channels (i.e., \emph{no} \conf~problem). This example shows that these two I/O requests can be serviced \orevII{completely} in parallel, which reduces the total service time of the two I/O requests. %

\vetII{To understand how much the \conf~problem can increase the total service time, we use the latency numbers for \emph{CMD}~\circled{1}, \emph{RD Operation}~\circled{2}, and \emph{Transfer}~\circled{3} from a performance-optimized SSD configuration (see \S\ref{sec:evaluation}). In a performance-optimized SSD configuration, \emph{CMD}, \emph{RD Operation}, and \emph{Transfer} take 10$n$s, 3$\mu$s, and 4$\mu$s, respectively~\cite{znand,cheong2018flash}. The total service time for the two read requests that experience the \conf~problem \orevIV{(as depicted in Figure \ref{fig:conflict} top)} is 11.01$\mu$s (i.e., $\emph{CMD}+\emph{RD Operation}+\emph{Transfer}+\emph{Transfer} = 11.01\mu s$). \orevIV{In contrast,} ideally \orevIV{(i.e., without a path conflict, as depicted in Figure \ref{fig:conflict} \orevV{bottom}), the total service time of the two requests is} 7.01$\mu$s (i.e., $\emph{CMD}+\emph{RD Operation}+\emph{Transfer}=7.01\mu s$). \orevIV{Thus, in} this \orevIV{simple} example, the \conf~problem \orevIV{increases the} average I/O access latency by 57\%, \orevIV{which in turn} results in lower SSD throughput. The performance overhead of \orevIV{\confs} can be even higher when (1) more than two I/O requests experience \orevIV{\confs}, and (2) the data transfer size of each request is \orevIV{larger} (e.g., a multi-plane operation; see \S\ref{sec:background}).} 

The \conf~problem affects the performance of read requests more than \orevIV{that of} write requests~\cite{gao2014exploiting,gao2017exploiting, jung2012physically, hu2011performance,park2010exploiting, hu2012exploring}. This is because data transfer time for \orevIV{a} read request %
is comparable to or longer than the flash read  latency\orevII{~\cite{znand,3Dxpoint,inteloptane}}, while the flash write latency \vetII{(e.g., 100$\mu$s for a performance-optimized SSD configuration~\cite{znand,cheong2018flash})} dominates the total service time \orevIV{of a} write request. 

We conclude that  the \conf~problem can significantly increase the total service time of I/O requests and limit SSD throughput, \vetII{especially for read-intensive workloads}. %

\vspace{-1mm}
\subsection{Approaches to Mitigate Path Conflicts \label{subsec:motivation_priorworks}}
We describe \orevII{two} major prior approaches to address \conf{} problem and their limitations.
\orevII{We quantitatively analyze the effectiveness of these approaches at mitigating path conflicts in \S\ref{subsec:motivation_effectiveness}.}

\head{Increasing Flash Channel Bandwidth} 
A recent work \orevIV{by} Kim et al.~\cite{kim2022networked} proposes the Packetized SSD (\pssd{}) (Figure ~\ref{fig:network}(b)), a technique to increase the flash channel bandwidth to 2$\times$ the channel bandwidth of the \baseline{}. This technique 1) utilizes control and data pins of the flash chip to transfer \orevII{\emph{both}} commands and data, thus increasing the channel bandwidth, and 2) integrates an on-die controller inside each flash chip \orevII{to enable packetization} between the flash controller and the flash chip. While \pssd{} can reduce the performance overhead
of \confs~by reducing the I/O transfer latency, \pssd{} imposes significant area overhead (i.e., 20\%~\cite{kim2022networked}) \orevII{in} each flash chip.

\head{Increasing Path Diversity}
Prior works~\cite{kim2022networked, tavakkol2014design, tavakkol2012network} propose techniques to mitigate \confs~by increasing the number of paths through which the SSD controller can access a flash chip (\orevII{i.e., these approaches increase }\emph{path diversity}). %

Kim et al.~\cite{kim2022networked} propose the Packetized Network SSD (\pnssd{}) (Figure~\ref{fig:network}(c)), a technique that provides two paths to access each flash chip, which reduces the performance overhead of \confs. \pnssd{} introduces an interconnection network similar to the 2D mesh topology~\cite{dally2004principles}, except in each dimension, flash chips are connected using a \orevII{shared bus}. As a result, an N$\times$N flash chip array has \emph{N} horizontal and \emph{N} vertical channels. In \pnssd{}, a flash chip can be accessed using either a horizontal channel or a vertical channel.

Tavakkol et al.~\cite{tavakkol2012network, tavakkol2014design} propose Network-on-SSD (\nossd{}) \orevII{(Figure \ref{fig:network}(d)),} which replaces the multi-channel shared bus architecture with a 2D mesh interconnection network of flash chips. \nossd{} significantly increases the path diversity compared to the \baseline{} and \pnssd{}. \orevII{However,} \nossd{} has two main weaknesses that limit its effectiveness at mitigating the \conf~problem. 
First, \nossd{} imposes significant
area \orevII{and} cost overhead due to \orevIV{1)} the integration of a buffered router (e.g., \orevII{with a} 16KB
buffer per router port) inside each flash chip\orevIV{, and 2) 4$\times$ increase in the number of I/O pins compared to a commodity flash chip}.
Second, \nossd{} does \emph{not} utilize the path diversity effectively as \nossd{} employs simple deterministic routing (i.e., the dimension-order routing algorithm~\cite{dally2004principles}) \orevII{that cannot adapt to the availability of multiple free paths between the flash controller and target flash chip}.

\vspace{-1mm}
\subsection{Effectiveness of Prior Approaches\label{subsec:motivation_effectiveness}}

\textbf{Methodology.} 
We study the effectiveness of prior approaches at mitigating \confs~using a state-of-the-art SSD simulator, \orevII{MQSim~\cite{tavakkol2018mqsim,mqsim-github}}, across nineteen real-world data-intensive workloads (see \S\ref{sec:evaluation} for our methodology). To this end, we measure the speedup of \pssd{}, \pnssd{}, and \nossd{} over the \baseline{} in a performance-optimized SSD configuration (see \S\ref{sec:evaluation}). We compare the speedup results with the speedup of the \orevII{ideal (i.e., path-conflict-free) SSD}. In the \ideal{}, we assume that each flash chip has a \orevII{\emph{direct separate channel}} to communicate with the SSD controller; therefore, no \conf~can happen. An I/O request does \emph{not} experience \confs~in the \ideal{}, but it can still be delayed if the target flash chip is busy.

\begin{figure}[!htb]
\centering
\includegraphics[width=\linewidth]{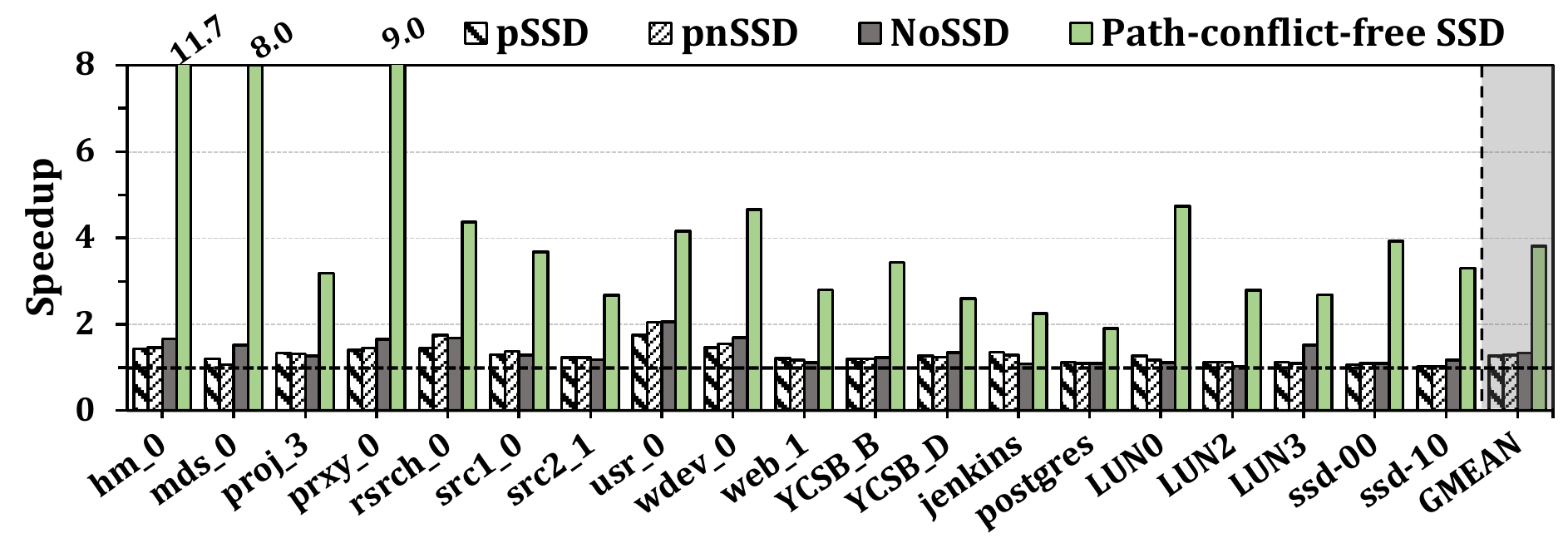}
\vspace{-6mm}
\caption{Performance of \pssd{}, \pnssd{}, \nossd{} and \orevII{the ideal} \ideal{} on a performance-optimized SSD configuration \orevII{(see \S \ref{sec:evaluation})}. Performance is shown in terms of speedup in overall execution time over \orevII{the} \baseline{}.}
\label{fig:motivation_baselines_performance}
\end{figure}

\head{Performance Results}
Figure ~\ref{fig:motivation_baselines_performance} shows the performance of \pssd{}, \pnssd{}, \nossd{} and \ideal{} compared to the \baseline{}. 
We make five \orevII{major} observations. 
First, the \orevII{\ideal{}} provides \vetII{an average of} 4$\times$ (\vetII{up to 11.74$\times$}) the performance of the \baseline{} %
since \orevII{it} does not \orevII{suffer from} any \orevII{\conf{}}. Second, \pssd{} shows an average performance improvement of 27\% over \baseline{} due to \orevII{its} increased channel bandwidth. 
Third, \pnssd{} provides an average performance improvement of 30\% over \baseline{} due to \orevII{its} increased path diversity.
Fourth, \nossd{} outperforms \baseline{} by 35\% on average due to the significantly increased path diversity provided by the interconnection network of flash chips. Fifth, although \nossd{} outperforms \pssd{} and \pnssd{}, \nossd{}'s speedup is still \orevII{greatly lower than the} \ideal{}'s speedup \orevII{(4$\times$)}. The main reason is that \nossd{} does not utilize the path diversity effectively.

We conclude that while prior approaches improve the performance of the SSD at a \orevII{large} cost overhead, none of \orevII{them effectively mitigate} the path conflict problem, \orevII{and a large potential remains between their performance and the performance of an SSD that does not suffer from path conflicts}.

\vspace{-1mm}
\subsection{Our Goal}
Based on our observations and analyses in \S\ref{subsec:motivation_pathconflict}, \S\ref{subsec:motivation_priorworks} and \S\ref{subsec:motivation_effectiveness}, we conclude that 1) the \conf~problem significantly limits the performance of modern SSDs, and  2) none of the prior approaches (i.e., \pssd{}, \pnssd{}, and \nossd{}) \orevII{effectively mitigate} the path conflict problem \orevII{even though they come with significant area \orevIV{overheads} and cost overheads}.%

\textbf{Our goal} is to fundamentally address the \conf~problem \orevII{in SSDs} by 1) providing path diversity \orevII{inside the SSD} at low cost, and 2) effectively utilizing the increased path diversity for communication between the SSD controller and flash chips. %

\section{\namePaper{}\label{sec:mechanism}}
\head{Overview} We \orevIV{design} \namePaper{}, a new mechanism that fundamentally addresses the \conf~problem in modern SSDs. \namePaper{} 1) provides rich path diversity between the SSD controller and flash chips by introducing a low-cost interconnection network of flash chips, and 2) utilizes the path diversity to identify \orevIV{and reserve} a conflict-free path for an I/O request .%
\namePaper{}'s design is based on three key techniques: (1) a low-cost interconnection network of flash chips in the SSD \orevIV{(\S \ref{subsec:ssd-interconnect})}, (2) reservation of a path between the flash controller and the flash chip for each I/O request \orevIV{(\S \ref{subsec:mechanism_path_reserve})}, and (3) a non-minimal fully-adaptive routing algorithm to utilize the path diversity provided by the interconnection network of flash chips \orevIV{(\S \ref{subsec:non_minimal_adaptive_routing})}. %

\subsection{Low-Cost Interconnection Network of \\Flash Chips}
\label{subsec:ssd-interconnect}
We want to provide rich path diversity between the SSD controller and flash chips at low cost. \namePaper{} can utilize the rich path diversity to eliminate \confs. To this end, we connect the flash chips using a \orevV{low-cost} interconnection network. The key design decision that enables our approach to be \orevIV{low cost} is \orevIV{the separation of} the router from the flash chip \orevIV{such that the flash chip is \orevIV{\emph{not}} modified}.

\begin{figure}[h]
\centering
\includegraphics[width=\linewidth]{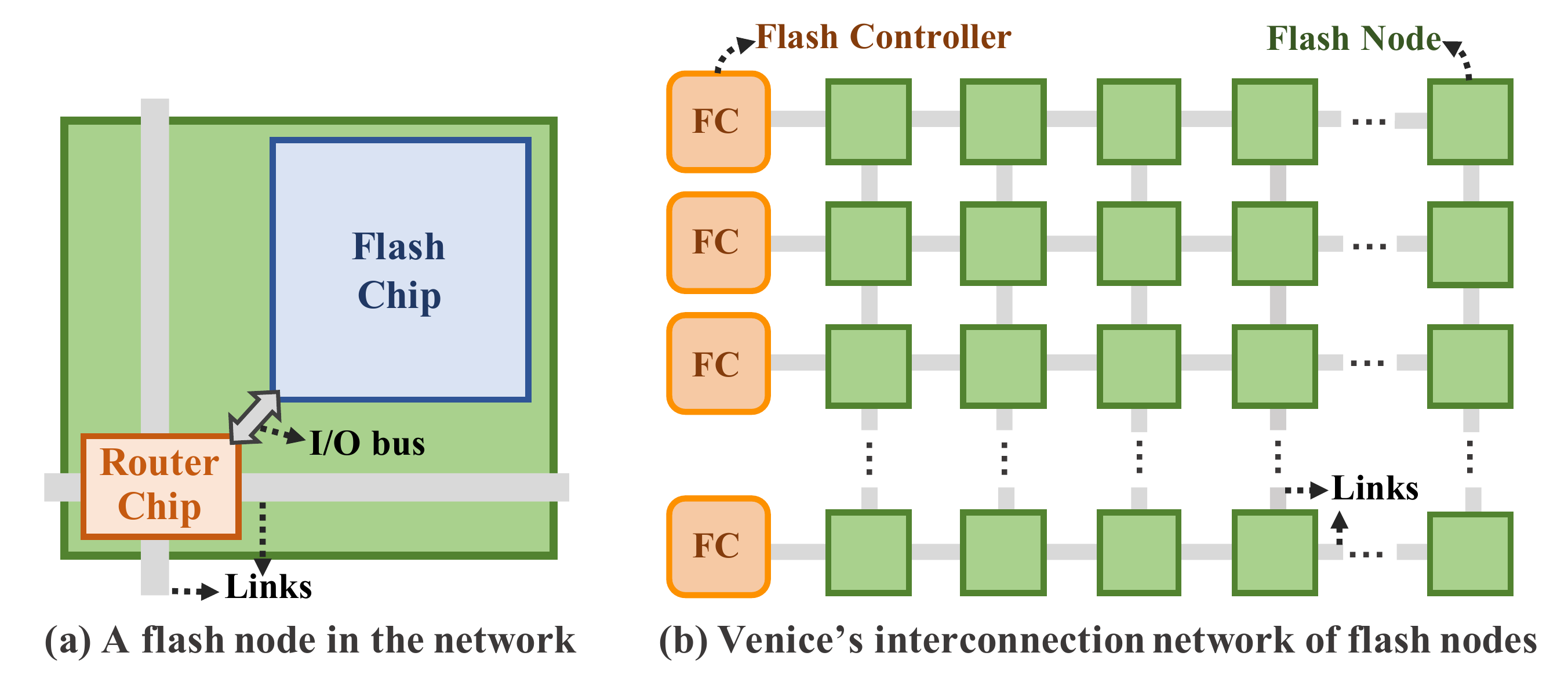}
\vspace{-3mm}
\caption{\namePaper{}'s low-cost \orevIV{interconnection network}}
\label{fig:flashnode}
\end{figure}

\orevIV{We introduce} a new \orevIV{building block}, called \emph{flash node}, which consists of a \emph{flash chip} and a separate \emph{router chip}. Figure~\ref{fig:flashnode}(a) shows \orevIV{a} flash node. In each flash node, a flash chip communicates with a router chip \orevIV{using its I/O data pins (i.e., injection/ejection \orevV{ports}) that are otherwise used for connecting \orevV{the} flash chip to the shared channel.}
\orevIV{Our design} connects the flash nodes using an interconnection network topology. Figure~\ref{fig:flashnode}(b) shows an example interconnection network of flash nodes using \orevIV{the} 2D mesh topology. The router chip in each flash node is connected to the router chips in the neighboring flash nodes using bidirectional links.

\subsection{Path Reservation\label{subsec:mechanism_path_reserve}}
\head{Key Idea} To ensure that the I/O request transfer does not experience path conflicts in the network, \namePaper{} reserves a conflict-free path between the flash controller and the target flash chip for each I/O request before starting the transfer. This technique \orevIV{avoids} the need for large buffers in each router that \orevIV{are} otherwise required to store the data of \orevIV{each} I/O request that experiences \orevV{a} \conf.

\begin{figure}[h]
\centering
\includegraphics[width=0.95\linewidth]{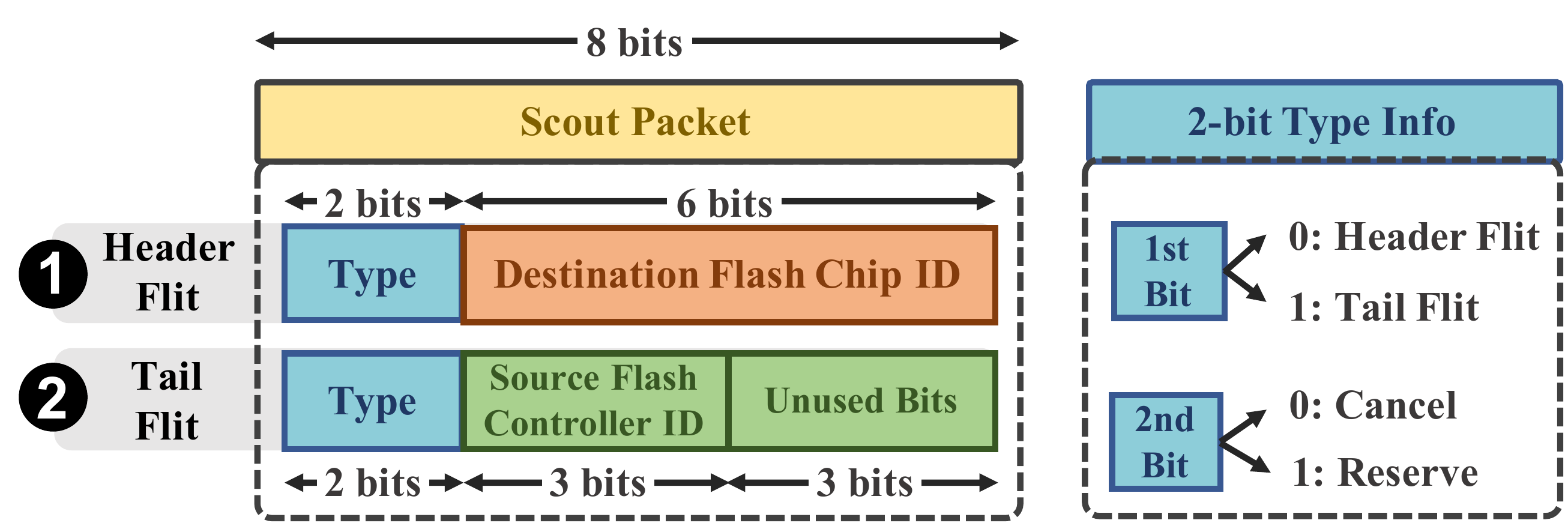}
\vspace{-3mm}
\caption{Structure of the \pkt{} for an SSD with 64 flash chips and 8 flash controllers}
\label{fig:packet}
\end{figure}

\head{Implementation} \namePaper{} identifies and reserves a path by sending a special packet called \orevIV{\emph{\pkt{}}}. Figure~\ref{fig:packet} shows the structure of a \pkt{} for an SSD with 64 flash chips and 8 flash controllers. The \pkt{} consists of two 8-bit scout flits, a \emph{header} flit~\circled{1}, and a \emph{tail} flit~\circled{2}. Each scout flit contains a 2-bit \emph{type} information, whose 1) most significant bit denotes whether the flit is the header flit or the tail flit, and 2) least significant bit denotes if the flit is in \textit{reserve} mode to reserve a link in the path or \textit{cancel} mode \orevIV{to cancel} a reservation. The destination flash chip ID is stored in the last 6 bits of the header flit (6 bits are required to represent 64 flash chips). In the tail flit, 3 bits are used to denote the source flash controller, and the other 3 bits are \orevIV{unused}. \orevV{The source flash controller ID is the same as the \pkt{} ID}.

\begin{figure}[h]
\centering
\includegraphics[width=0.95\linewidth]{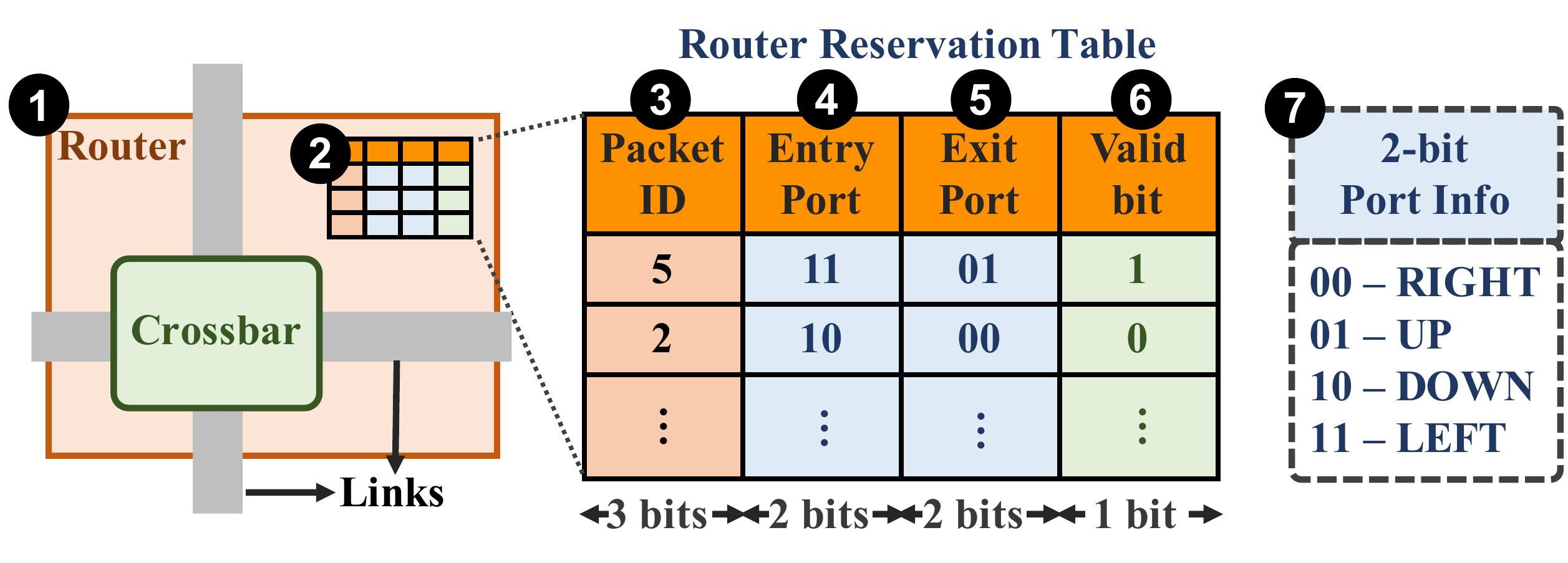}
\vspace{-3mm}
\caption{\vetIV{Structure of a router in \namePaper{}'s interconnection network of flash nodes (assuming 8 flash controllers)}}
\label{fig:router}
\end{figure}

For \orevIV{a given} I/O request, \namePaper{} checks if the closest flash controller to the target flash chip is available. If so, \namePaper{} selects the flash controller to handle the I/O request. Otherwise, %
\namePaper{} uses the nearest free flash controller.
The source flash controller sends a \pkt{} \orevIV{in} \textit{reserve} mode to identify and reserve a path to the destination chip. 
\namePaper{} uses a routing algorithm (e.g., \orevV{the} non-minimal fully-adaptive routing algorithm as described in \S\ref{subsec:non_minimal_adaptive_routing}) to route a \pkt{} from the source flash controller to the destination flash chip. \namePaper{} reserves the interconnection network's links that a \pkt{} takes to reach the destination node. \orevIV{Each reserved link is bidirectional, which enables data transfer 1) from the flash controller to the flash chip (e.g., a write request) using the \emph{forward path}, and 2) from the flash chip to the flash controller (e.g., a read request) using the \emph{backward path}.} To this end, \orevIV{we introduce} a table, called \emph{router \orevIV{reservation} table}, to each router chip.
\orevV{Figure~\ref{fig:router} shows the structure of \namePaper{}'s router~\circled{1} and router reservation table}~\circled{2}.
\orevV{The router reservation table} keeps track of \orevIV{1) \orevV{the packet ID~\circled{3}, which is the same as the source flash controller ID from which the packet was sent}, and 2)} which \orevIV{two ports (i.e., entry~\circled{4} and exit~\circled{5} ports) are connected bidirectionally} 
\orevIV{(based on reservations \orevVI{that were made})}. \vetIV{Each row in the router reservation table has a \emph{valid} bit~\circled{6} that shows whether the entry is valid.} 
\vetIV{The packet ID has \emph{log(n)} bits to denote one of the \emph{n} flash controllers, which allows up to \emph{n} \pkts{} to be sent simultaneously. In our example interconnection network configuration with 8 flash controllers, we need 3 bits for the packet ID.} \orevV{The entry port~\circled{4} and exit port~\circled{5} information in the router reservation table \orevVI{each contains} 2 bits~\circled{7} to denote one of the four ports in the router.}

\orevIV{When the \pkt{} arrives at the destination flash chip, \namePaper{} has already reserved the conflict-free \emph{forward} and \emph{backward} paths.}
\orevIV{The router connected to the destination flash chip uses the \emph{backward} path to send} the \pkt{} back to the source flash controller. 
Once the source flash controller receives \orevIV{back} the \pkt{}, it schedules the I/O request transfer using the reserved path.

If a \pkt{} is unable to find a free link at a router during the path reservation \orevIV{process}, the router enables the \textit{cancel} mode in the \pkt{}, which cancels the reservation by removing its entry in the router \orevIV{reservation} table. The \pkt{} backtracks along its path to a previously traversed router (i.e., upstream router). Depending on the routing algorithm's adaptivity, the \pkt{} may either try a different free output link in the upstream router or backtrack further (i.e., to the upstream router of the upstream router). 
In case the \pkt{} is unable to find a free output link during backtracking, the \pkt{} can arrive back \orevV{at} the flash controller \emph{without} reserving a path.
\orevIV{When the source flash controller receives the \pkt{} in \emph{cancel} mode, it retries the path reservation process \orevV{immediately} by sending a new \pkt{}.\footnote{\orevV{We study more optimizations, including when to resend the \pkt{}, in the path reservation process in \orevVI{the} extended version of our paper~\cite{nadig2023venice}.}}}

\vspace{-1mm}
\subsection{Utilizing Path Diversity\label{subsec:non_minimal_adaptive_routing}} %

\head{Key Idea} 
To effectively utilize the interconnection network's path diversity,
\namePaper{} uses a \emph{non-minimal fully-adaptive} routing algorithm for routing a \pkt{} (\vetIV{during the path reservation process}) through the interconnection network of flash nodes. 
\namePaper{}'s non-minimal fully-adaptive routing algorithm dynamically identifies a conflict-free path between the flash controller and the flash chip. This algorithm effectively utilizes the idle links in the interconnection network to find a non-minimal path when a minimal path is unavailable. 

\begin{figure}[!htb]
\centering
\includegraphics[width=0.95\linewidth]{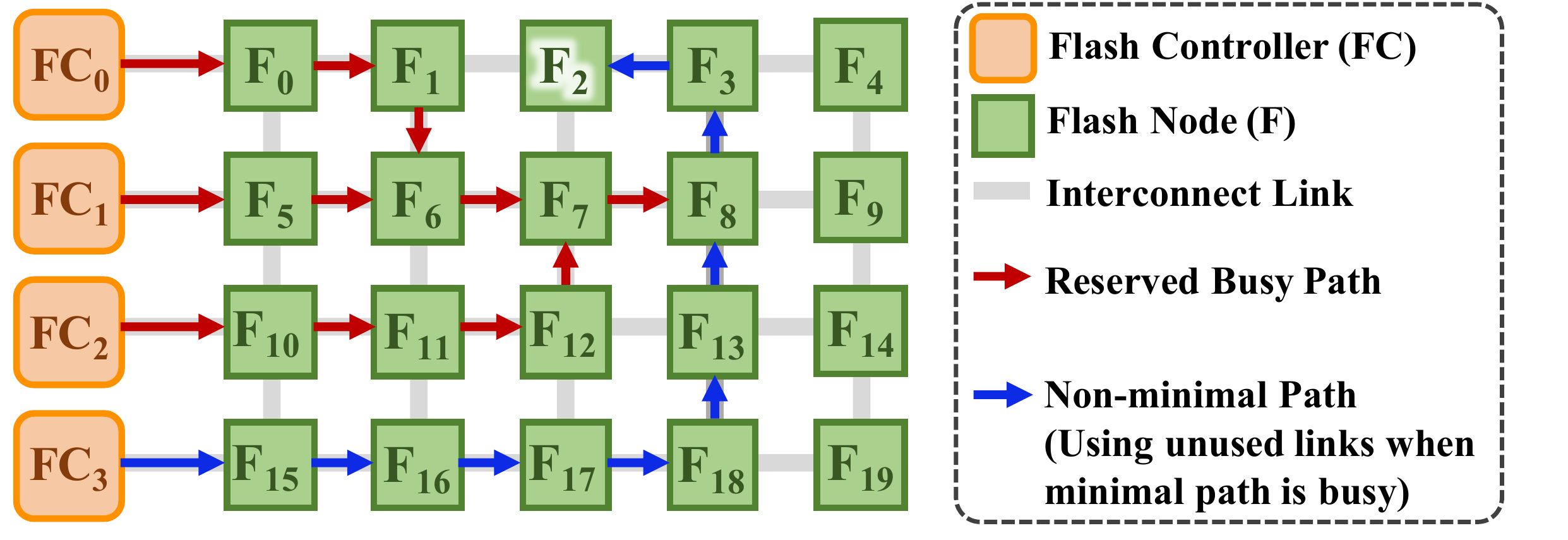}
\vspace{-3mm}
\caption{\vetIV{Example demonstrating \orevV{how \namePaper{}'s} non-minimal routing algorithm \orevV{finds} a conflict-free path in the interconnection network of flash nodes}}
\label{fig:nonminimal}
\end{figure}

\orevIV{Figure \ref{fig:nonminimal} illustrates} how a non-minimal fully-adaptive routing algorithm helps to mitigate the \conf~problem \orevIV{via} an example. 
In this example, 
a new I/O request $R$ has $F_2$ as its destination flash chip.
In the network, there are three paths already reserved for other I/O requests (marked in red in Figure~\ref{fig:nonminimal}): \orevIV{$FC_0 \rightarrow F_0 \rightarrow F_1 \rightarrow F_6$, $FC_1 \rightarrow F_5 \rightarrow F_6 \rightarrow F_7 \rightarrow F_8$, and $FC_2 \rightarrow F_{10} \rightarrow F_{11} \rightarrow F_{12} \rightarrow F_7$.}
\vetIV{The \emph{only} free flash controller, $FC_3$, is assigned to request $R$. Each minimal path from $FC_3$ to $F_2$ has at least one \emph{busy} link, and thus, is \emph{not} \orevV{path-}conflict-free. }
However, \vetIV{there are a number of non-minimal paths from the $FC_3$ to $F_2$ that are \orevV{path-}conflict-free. An example is $FC_3 \rightarrow F_{15} \rightarrow F_{16} \rightarrow F_{17} \rightarrow F_{18} \rightarrow F_{13} \rightarrow F_{8} \rightarrow F_3 \rightarrow F_2$ (shown in \orevIV{blue} in Figure~\ref{fig:nonminimal}). \namePaper{} \orevVI{uses} a non-minimal fully-adaptive routing algorithm \orevV{(described in Algorithm \ref{alg:nonminimal})} during the path reservation process to increase its ability to mitigate the \conf~problem.}

\vetIV{To effectively use a non-minimal fully-adaptive routing algorithm during the path reservation process in \namePaper{}, we should address two key challenges\orevV{:}} 
1) performance overhead of \orevIV{exercising a} non-minimal path \orevIV{to service the I/O request}, and 2) %
\orevIV{the need to avoid}  deadlock/livelock. %
\orevIV{We describe these two key challenges and \namePaper{}'s techniques to address them.}

\noindent\textbf{\orevIV{Performance} Overhead of \orevIV{Exercising} a Non-Minimal Path.} 
The increased path length of the non-minimal route \vetIV{can cause two issues.} First, a non-minimal path may lead to an increase in the command/data transfer time and \orevIV{thus the} overall latency for \orevIV{servicing} the I/O request. After reserving a free path, the transfer latency ($T_{transfer}$) in \vetIV{seconds} can be calculated using Equation~\ref{eq:latency}:
\vspace{-0.1cm}
\begin{equation}
\label{eq:latency}
T_{transfer} = [distance + (transfer_{size} / link\_width)] \times link_{lat.}
\end{equation}
\noindent where \emph{distance},  \emph{$transfer_{size}$}, \emph{link\_width}, and \emph{$link_{lat.}$} are the number of links between the flash controller and flash chip (i.e., hops), the command/data transfer size in terms of the number of bytes, the link width in terms of the number of bytes,  and \orevV{the latency of a single transfer (of size link\_width) on the link \vetIV{in seconds}},  respectively. \vetIV{A non-minimal path has a longer \emph{distance} compared to a minimal path. \orevV{We study the latency overhead of a non-minimal path for I/O data and flash command transfer.} For the I/O data transfer, the \orevIV{performance} overhead of a longer path is negligible. This is because \orevV{$transfer_{size}$ dominates} $T_{transfer}$ as I/O data transfers are} 
large in size (e.g., 16KB). 
\orevV{Flash commands, on the other hand, have a small $transfer_{size}$ (\orevVI{i.e., only} a few bytes), and thus, a longer path can significantly increase their transfer latency. 
As discussed in \S\ref{subsec:motivation_pathconflict}, the service time of an I/O request consists of flash command transfer time, flash operation latency and I/O data transfer time. The total service time is dominated by the I/O data transfer time and the flash operation latency, and thus, the longer command transfer time due to a non-minimal path has a negligible effect on the total service time of the I/O request.} 

\vetIV{Second, a non-minimal path occupies more links in the interconnection network compared to a minimal path. If a minimal path is used instead of a non-minimal path, the extra links of a non-minimal path can potentially be used for transferring other ongoing requests, which increases the effectiveness of \namePaper{}. To this end, \namePaper{} attempts to find \orevV{path-}conflict-free minimal paths during the path reservation process as much as possible \orevV{(as described in Algorithm \ref{alg:nonminimal})}.}

\head{\orevIV{Need to Avoid} Deadlock/Livelock} 
\orevIV{A} non-minimal fully-adaptive routing algorithm can potentially cause (1) \vetIV{deadlock in the interconnection network\orevVI{~\cite{duato2003interconnection,dally2004principles,moscibroda2009case,gravano1994adaptive,fallin2011chipper, fallin2012hird,glass1992turn, fu2011abacus,shafiee2011application,ebrahimi2017ebda,duato1995necessary,ausavarungnirun2014design}}, where multiple network packets cannot move forward as they circularly depend on each other to free up resources (e.g., channels, buffers)}, 
and (2) livelock\orevVI{~\cite{duato2003interconnection,dally2004principles,moscibroda2009case,gravano1994adaptive,fallin2011chipper, fallin2012hird,navaridas2009understanding,berman1992adaptive,glass1992turn,coli1995adaptive,ausavarungnirun2014design}}, where \orevV{at least one packet keeps} traversing the network without reaching \orevV{its} destination. \vetIV{\namePaper{}'s interconnection network can experience deadlock and livelock \orevV{only} during the path reservation process where the \pkts{} are routed using the non-minimal fully-adaptive routing algorithm.\footnote{\orevV{Once a path between the flash controller and the destination flash chip is reserved, there is no 1) deadlock\orevVI{,} as there is no \conf~during I/O data or flash command transfer, or 2) livelock\orevVI{,} as the path from the source flash controller and the destination flash chip is \orevVI{deterministically set as a circuit}.}}}

\namePaper{} handles deadlock \orevIV{by} using backtracking of a \pkt{}. When a \pkt{} experiences path conflict \orevIV{during the path reservation process}, it backtracks along its path to the previously visited router (\vetIV{i.e., the upstream router})  in order to choose a different path. %
As a result, a \pkt{} is \orevIV{never} blocked \orevV{due to resource unavailability} in the network and \vetIV{deadlock does not happen.}

\namePaper{} handles livelock by restricting the number of times a \pkt{} can visit the same router. A \pkt{} can \orevIV{reserve} each output port of a router only once
and hence, the \pkt{} may revisit the same router \orevIV{at most} three times\footnote{ \orevVI{The number of times a \pkt{} can revisit a router is four minus one, i.e., number of ports in a router minus the entry port of the \pkt{}.}} in an interconnection network with \orevIV{a} 2D mesh topology.
\orevIV{When a \pkt{} revisits the same router three times, the \pkt{} traces its path back to the upstream router %
(using the router reservation table) and attempts to reserve a different output port in the upstream router. %
In the worst case, when a \pkt{} fails to reserve a path to the destination after visiting all the routers at most three times, it will return to the source flash controller.} \vetIV{The flash controller \orevVI{immediately} sends a new \pkt{} to retry the path reservation.}

\noindent\textbf{Implementation.} \vetIV{Algorithm~\ref{alg:nonminimal} shows the pseudocode for the non-minimal fully-adaptive routing algorithm in \namePaper{}. The inputs to the algorithm are\orevV{:} 1) \pkt{} ID, 2) current router ID, 3) \pkt{}'s destination router ID, 4) input port \orevV{of the router through} which the \pkt{} has arrived, 5) \orevV{the status (free or busy) of the output ports in the router}, and 6) the interconnection network structure in terms of number of rows and columns \orevV{(assuming a 2D mesh topology)}. The algorithm returns the output port in which the \pkt{} should traverse to the downstream router.} 

\begin{algorithm}[h]
\scriptsize
\setstretch{0.9}
\caption{\small \textbf{\namePaper{}'s Non-Minimal Fully-Adaptive Routing Alg.}}
\label{alg:nonminimal}
\centering
\textbf{Input:} \pkt{} ID: $P_{ID}$, %
    current router ID: $ID_{rc}$, %
    \pkt{}'s destination router ID: $ID_{rd}$, %
    the Input\_port, the output ports' status, %
    and network structure: $N_{r}$ rows and $N_{c}$ columns \\
\textbf{Output:} Output\_port

\begin{algorithmic}[1]

\Procedure{Find output port}{}
      \State  $Diff_x = ID_{rd}\%N_{c} - ID_{rc}\%N_{c}$
      \State  $Diff_y = ID_{rd}/N_{c} - ID_{rc}/N_{c}$
      \State $Output_{list}.clear()$
      \State \textbf{Switch}($Diff_x$ and $Diff_y$) \algcomment{Nine cases in total as $Diff_x$ and $Diff_y$ can \orevV{each} be a positive, zero, or negative value}
      \State \hspace{0.3cm} $Case_1$: $Diff_x > 0$ \& $Diff_y > 0$
      \State \hspace{0.6cm} 
      \textbf{if}($Right.status() == free$) \textbf{then}
        \State \hspace{0.9cm} $Output_{list}.add(Right)$
      \State \hspace{0.6cm}
      \textbf{if}($Up.status() == free$) \textbf{then}
        \State \hspace{0.9cm} $Output_{list}.add(Up)$
      \State \hspace{0.3cm} $Case_2$: $Diff_x > 0$ \& $Diff_y < 0$
      \State \hspace{0.6cm} 
      \textbf{if}($Right.status() == free$) \textbf{then}
        \State \hspace{0.9cm} $Output_{list}.add(Right)$
      \State \hspace{0.6cm}
      \textbf{if}($Down.status() == free$) \textbf{then}
        \State \hspace{0.9cm} $Output_{list}.add(Down)$
        \State \hspace{0.3cm} $Case_3$: $Diff_x > 0$ \& $Diff_y == 0$
      \State \hspace{0.6cm} 
      \textbf{if}($Right.status() == free$) \textbf{then}
        \State \hspace{0.9cm} $Output_{list}.add(Right)$
    \State \hspace{0.3cm} $Case_4$: $Diff_x < 0$ \& $Diff_y > 0$ ...
    \State \hspace{0.3cm} $Case_5$: $Diff_x < 0$ \& $Diff_y < 0$ ...
    \State \hspace{0.3cm} $Case_6$: $Diff_x < 0$ \& $Diff_y == 0$ ...
    \State \hspace{0.3cm} $Case_7$: $Diff_x == 0$ \& $Diff_y > 0$ ...
    \State \hspace{0.3cm} $Case_8$: $Diff_x == 0$ \& $Diff_y < 0$ ...
    \State \hspace{0.3cm} $Case_9$: $Diff_x == 0$ \& $Diff_y == 0$
    \State \hspace{0.6cm} $Output_{list}.add(Ejection)$
    \State \textbf{end}

\algcomment{check the number of output ports in the output list}
\State \hspace{0.0cm} \textbf{if} ($Output_{list}$.size() == 2)  \textbf{then}
\State \hspace{0.3cm} Output\_port = randomly select one output port from $Output_{list}$
\State \hspace{0.3cm} Routing\_Reservation\_Table.insert ($P_{ID}$, Input\_port, Output\_port)
\State \hspace{0.0cm} \textbf{else if} ($Output_{list}$.size() == 1)  \textbf{then}
\State \hspace{0.3cm} Output\_port = $Output_{list}.top()$
\State \hspace{0.3cm} Routing\_Reservation\_Table.insert ($P_{ID}$, Input\_port, Output\_port)
\State \hspace{0.0cm} \textbf{else} 
\State \hspace{0.3cm} $Non\_minimal\_Output_{list}$.clear();
\State \hspace{0.3cm} \textbf{if} (Up.status() == free \& Up != Input\_link)  \textbf{then}
\State \hspace{0.6cm} $Non\_minimal\_Output_{list}$.add(Up)
\State \hspace{0.3cm} \textbf{if} (Down.status() == free \& Down != Input\_link)  \textbf{then}
\State \hspace{0.6cm} $Non\_minimal\_Output_{list}$.add(Down)
\State \hspace{0.3cm} \textbf{if} (Right.status() == free \& Right != Input\_link)  \textbf{then}
\State \hspace{0.6cm} $Non\_minimal\_Output_{list}$.add(Right)
\State \hspace{0.3cm} \textbf{if} (Left.status() == free \& Left != Input\_link)  \textbf{then}
\State \hspace{0.6cm} $Non\_minimal\_Output_{list}$.add(Left)  \vspace{0.2cm}
\State \hspace{0.3cm} \textbf{if} ($Non\_minimal\_Output_{list}$.size() > 0) \textbf{then}
\State \hspace{0.6cm} Output\_port = randomly select one output port from $Non\_minimal\_Output_{list}$
\State \hspace{0.6cm} Routing\_Reservation\_Table.insert ($P_{ID}$, Input\_port, Output\_port)
\State \hspace{0.3cm} \textbf{else}
\State \hspace{0.6cm} Output\_port = Input\_port \algcomment{traverse back to the upstream router}
\State \textbf{end}
\EndProcedure
\end{algorithmic}
\end{algorithm}

\vetIV{To find an appropriate output port, the algorithm first attempts to find a free output port that leads to a minimal path \orevV{(lines 2-32)}. To this end, the algorithm compares the current router ID with the \pkt{}'s destination router ID in both \emph{X} (horizontal) and \emph{Y} (vertical) dimensions. Based on the comparison, it switches among nine cases \orevV{(lines 5-26)}. In each case, the algorithm checks the status of the corresponding output port and adds the output port to the output list if the corresponding output port is free (e.g., \orevV{lines 6-10}).} 

\vetIV{The algorithm checks the number of output ports added to the output list \orevV{(line 27)}. In a 2D mesh topology, the size of the output list can be either two, one, or zero. If there are \emph{two} output port candidates in the output list, the algorithm randomly selects one output port using a pseudo-random number generator. We use a simple 2-bit Linear-Feedback Shift Register (LFSR)~\cite{wang1988linear} present in the router for the pseudo-random number generation. The algorithm adds an entry to the router reservation table using the \pkt{} ID, the input port, and the selected output port. The \pkt{} \orevV{then proceeds} to the downstream router using the selected output port \orevV{(lines 27-29)}. If there is only \emph{one} output port candidate in the output list, the algorithm selects that output port and \orevV{records it in} the router reservation table \orevV{(lines 30-32)}.}

\vetIV{However, if the output list is empty, the algorithm has failed to find any free output port that leads to a minimal path. In this case, the algorithm \emph{misroutes} the \pkt{} via a free output port that leads to a non-minimal path. To this end, the algorithm randomly selects any free output port (except the ejection port) and adds an entry to the router reservation table \orevV{(lines 34-45)}. 
If the only available free output port is the port that results in backtracking to the upstream router, the \pkt{} \orevV{travels} back to the upstream router, and the algorithm does \emph{not} reserve the selected output port \orevV{(lines 46-47)}. When the upstream router receives the backtracking \pkt{}, it clears the reservation entry for the \pkt{} in the router reservation table and tries another available output port, if any.}
\orevV{This algorithm is used in conjunction with the livelock avoidance mechanism of \namePaper{} that is described earlier in this section (not shown in Algorithm \ref{alg:nonminimal}). Note that it is also possible to employ other non-minimal fully-adaptive routing algorithms in \namePaper{} instead of this specific one we use and evaluate.}

\section{Methodology\label{sec:evaluation}}

\head{Simulation Methodology\label{subsec:eval_setup}}
We evaluate \namePaper{} using MQSim~\cite{tavakkol2018mqsim,mqsim-github}, a state-of-the-art \orevII{open-source} SSD simulator.
MQSim models all components of the SSD, including host interface, SSD controllers, flash controllers, and flash chips. MQSim supports multi-queue SSDs and measures the end-to-end latency~\cite{tavakkol2018mqsim}, which makes it a suitable tool for our study. We model two SSD configurations:
1) a \orevII{\emph{performance-optimized}} configuration based on Samsung Z-NAND SSD~\cite{znand, cheong2018flash} and 2) a \orevII{\emph{cost-optimized}} configuration based on Samsung PM9A3 SSD~\cite{pm9a3}. \tab{\ref{tab:config}} provides details of the storage characteristics of the two configurations and \namePaper{}'s design parameters used in our evaluation. 
\begin{table}[ht]
    \caption{Evaluated \orevII{configurations} \& Venice \orevII{parameters}}
    \vspace{-0.3cm}
    \centering
    \resizebox{\columnwidth}{!}{
    \begin{tabular}{|@{} c||l @{}|}
    \hline
    \multirow{7}{*}{
    \begin{tabular}{c}
    \textbf{Performance-optimized} \\ \textbf{SSD~\cite{znand,cheong2018flash}}
    \end{tabular}
    }
    & 240GB, Z-NAND~\cite{znand,cheong2018flash,SZ985}, \\ & 8-GB/s External I/O bandwidth (4-lane PCIe Gen4);\\
    & 1.2-GB/s Flash Channel I/O rate\\\cline{2-2}
    & \textbf{NAND Config}: 8 channels, 8 chips/channel, \\ & 1 die/chip, 2 planes/die, 128Gb die capacity, \\ &  1024 blocks/plane, 768 pages/block, 4KB page \\\cline{2-2}
    & \textbf{Latencies}: Read(\tr): 3$\mu$s; Erase (\tbers): 1$m$s\\ &  Program (\tprog): 100$\mu$s  \\
    \hline
    \hline
    \multirow{7}{*}{
    \begin{tabular}{c}
    \textbf{Cost-optimized} \\ \textbf{SSD~\cite{pm9a3}}
    \end{tabular}
    }
    & 1TB, 3D TLC NAND Flash, \\ & 8-GB/s External I/O bandwidth (4-lane PCIe Gen4);\\
    & 1.2-GB/s Flash Channel I/O rate\\\cline{2-2}
    & \textbf{NAND Config}: 8 channels, 8 chips/channel, \\ & 1 die/chip, \orevII{2} planes/die, \orevII{1024} blocks/die, 16KB page \\\cline{2-2}
    & \textbf{Latencies}: Read (\tr): 45$\mu$s; Erase (\tbers): 3.5$m$s\\ & Program (\tprog): 650$\mu$s \\
    \hline
    \hline
    \multirow{6}{*}{
    \begin{tabular}{c}
    \textbf{Venice Design Parameters}
    \end{tabular}
    }
    & \textbf{Topology.} 8$\times$8 2D mesh topology, 8-bit 1 GHz links, \\ & \reva{One router next to each flash chip} \\
    & \textbf{Router Architecture.} Two 8-bit buffers per port, \\ & \reva{1 GHz frequency} \\
    & \textbf{Routing Algorithm.} Non-minimal fully-adaptive \\
    & \textbf{Switching.} Circuit switching~\cite{dally2004principles} \\
    \hline
   \end{tabular}
   }
   \label{tab:config}
\end{table}
\orevII{To evaluate \namePaper{}'s power overhead, we measure the power consumption of each router and network link in the interconnection network (see \S \ref{subsec:overhead}). We analyze 1) the average power consumption of a router by implementing its hardware description language (HDL) model and synthesizing it for the UMC 65nm technology node~\cite{umc}, and 2) the average power consumption for a 4KB data transfer over each network link using \orevIV{the} ORION 3.0 \cite{kahng2015orion3} power model.
To model the power consumption \orevIV{of} flash read and write operations, we use the power values from Samsung Z-SSD SZ985~\cite{SZ985}.  
}

\begin{figure*}[b]
\centering
\includegraphics[width=0.95\textwidth]{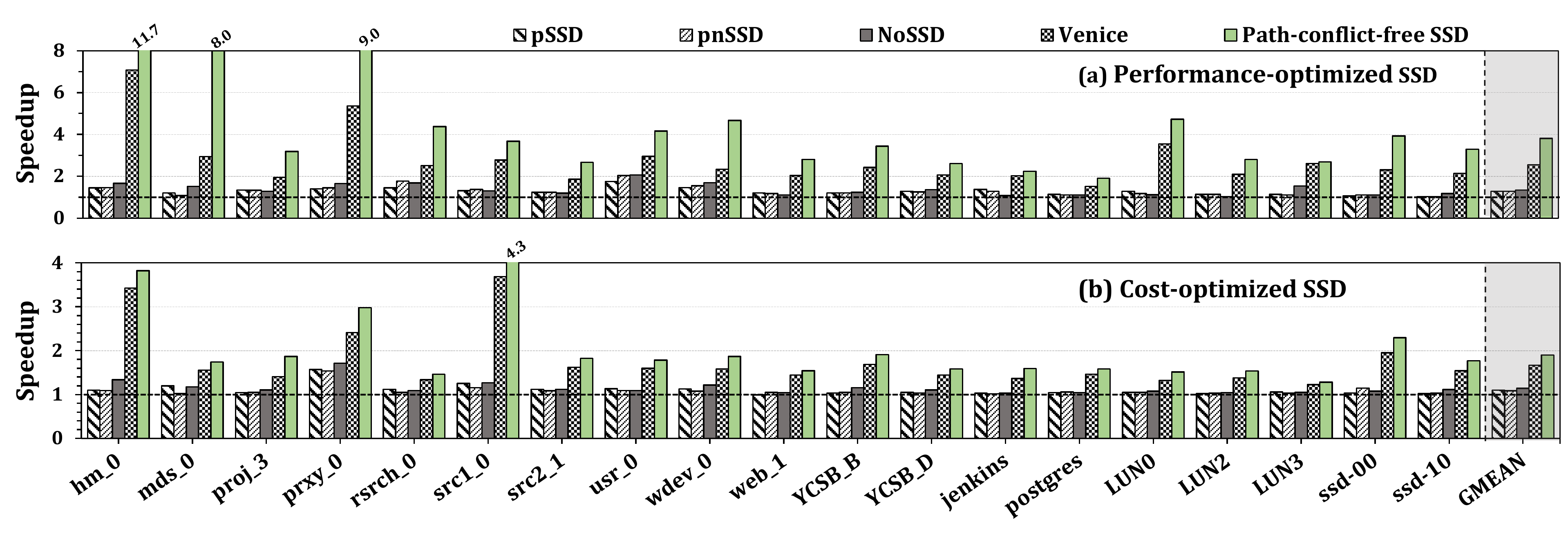}
\vspace{-3mm}
\caption{Performance of \pssd{}, \pnssd{}, \nossd{}, \namePaper{} and \ideal{} on performance-optimized \orevII{(top)} and cost-optimized \orevII{(bottom)} \orevIV{SSD} configurations. Performance is shown in terms of speedup in overall execution time over \baseline{}.}
\label{fig:results_overall_speedup}
\end{figure*}

\head{Evaluated Systems}
We compare \namePaper{} with the following prior approaches \orevII{(described in \S \ref{sec:motivation}):} (1) \Baseline{}, a typical SSD with multi-channel shared bus architecture; (2) Packetized SSD (\pssd{})~\cite{kim2022networked}, a prior proposal that uses packetization to \orevII{double the} flash channel bandwidth; (3) Packetized Network SSD (\pnssd{})~\cite{kim2022networked}, a technique that increases path diversity by introducing vertical flash channels; 
(4) NoSSD~\cite{tavakkol2012network, tavakkol2014design}, a state-of-the-art proposal on interconnection network of flash chips that uses a deterministic minimal routing policy to route I/O requests. We compare \namePaper{} and the four prior approaches with \orevII{an ideal} \ideal{}. In a \ideal{}, \orevII{we assume that} each flash chip \orevII{has a \emph{direct separate channel} to communicate with the SSD \orevIV{controller}}, which eliminates path conflicts.

\head{Workloads\label{subsec:eval_workloads}}
We select \orevII{nineteen data-intensive} storage \orevII{workloads} from MSR Cambridge traces~\cite{narayanan2008write}, Yahoo! Cloud Serving Benchmark (YCSB) suite~\cite{cooper2010benchmarking}, Slacker~\cite{harter2016slacker}, SYSTOR '17~\cite{lee2017understanding} and YCSB RocksDB traces~\cite{yadgar2021ssd} that are collected from real enterprise and datacenter workloads. 
These workloads \orevII{are chosen to represent diverse I/O access patterns}, \orevII{with different read and write ratios}, I/O request sizes, and inter-request arrival times. 
Table~\ref{table:trace} reports the characteristics of the workloads chosen for our evaluation. 

\begin{table}[h]
    \centering
    \scriptsize
    \caption{Characteristics of the \orevII{evaluated} I/O traces}
    \label{tab:workloads}
    \vspace{-0.3cm}
    \resizebox{\linewidth}{!}{
    	\begin{tabular}{|c||c|c|c|c|}\hline
    	\textbf{} &	\textbf{Traces} & \textbf{Read} \% &
            \begin{tabular}{c}    
            \textbf{Avg. Request} \\\textbf{Size (KB)}
            \end{tabular}
             &
             \begin{tabular}{c}    
            \textbf{Avg. Inter-request} \\\textbf{Arrival Time ($\mu$s)}
            \end{tabular}
              \\
                \hline
                \hline
                \multirow{10}{*}{\textbf{MSR Cambridge~\cite{narayanan2008write}}}
    		& hm\_0 & 36 & 8.8 & 58\\\cline{2-5}%
    		  & mds\_0 & 12 & 9.6 & 268\\\cline{2-5}%
    		& proj\_3 & 95 & 9.6 & 19\\\cline{2-5}
    		& prxy\_0 & 3 & 7.2 & 242\\\cline{2-5}
    		& rsrch\_0 & 9 & 9.6 & 129\\\cline{2-5}
    		& src1\_0 & 56 & 43.2 & 49\\\cline{2-5}
    		& src2\_1 & 98 & 59.2 & 50\\\cline{2-5}
    		& usr\_0 & 40 & 22.8 & 98\\\cline{2-5}
    		& wdev\_0 & 20 & 9.2 & 162\\\cline{2-5}
    		& web\_1~ & 54 & 29.6 & 67\\\hline
                \multirow{2}{*}{\textbf{YCSB~\cite{cooper2010benchmarking}}}
    		& YCSB\_B & 99 & 65.7 & 13\\\cline{2-5}
    		& YCSB\_D & 99 & 62 & 14\\\hline
                \multirow{2}{*}             {\reva{\textbf{Slacker~\cite{harter2016slacker}}}}
                & \reva{jenkins} & 94 & 33.4 & 615\\\cline{2-5}
    		& \reva{postgres} & 82 & 13.3 & 382\\\hline
                \multirow{3}{*}{\reva{\textbf{SYSTOR '17~\cite{lee2017understanding}}}}
                & \reva{LUN0} & 76 & 20.4 & 218\\\cline{2-5}
    		& \reva{LUN2} & 73 & 16 & 320\\\cline{2-5}
                & \reva{LUN3} & 7 & 7.7 & 3127\\\hline
                \multirow{2}{*}{\reva{\textbf{YCSB RocksDB~\cite{yadgar2021ssd}}}}
                & \reva{ssd-00} & 91 & 90 & 5\\\cline{2-5}
    		& \reva{ssd-10} & 99 & 11.5 & 2\\\hline 
    	\end{tabular} 
	}
	\label{table:trace}
\end{table}

To evaluate \namePaper{} under real-world scenarios, where multiple workloads access the same SSD, we create \vetII{\emph{mixed}} workloads by combining two or three independent storage workloads. Table~\ref{tab:mixed_workloads} shows six mixed workloads and their different characteristics. 

Mixed workloads \orevIV{usually} have a higher intensity of I/O requests (i.e., \orevII{lower} inter-request arrival time between I/O requests), which \orevII{likely exacerbates} the \conf~problem in the SSD.
\begin{table}[h]
 \caption{Characteristics of mixed workloads}
 \vspace{-3mm}
    \label{tab:mixed_workloads}
\centering
\normalsize
 \setstretch{1}
   \renewcommand{\arraystretch}{1} %
\setlength{\tabcolsep}{2pt}
  \resizebox{\linewidth}{!}{%
\begin{tabular}{|c||c|c|c|}
\hline
\textbf{Mix} & 
\begin{tabular}{c}    
\textbf{Constituent} \\\textbf{Workloads~\cite{narayanan2008write,cooper2010benchmarking}}
\end{tabular} 
& \textbf{Description} &  
\begin{tabular}{c}    
\textbf{Avg. Inter-request} \\\textbf{Arrival Time ($\mu$s)}
\end{tabular} \\
\hline
\hline
\textbf{mix1} &  
src2\_1 and proj\_3 & 
\begin{tabular}{l}    
 Both workloads are read-intensive
\end{tabular}
& 5.8
\\
\hline
\textbf{mix2}                      &
\begin{tabular}{l}
src2\_1, proj\_3 and YCSB\_D 
\end{tabular} &
\begin{tabular}{l} 
All three workloads are read-intensive
\end{tabular}
& 8.4
\\
\hline
\textbf{mix3} &  
prxy\_0 and rsrch\_0 & 
\begin{tabular}{l}
Both workloads are write-intensive
\end{tabular}
& 93
\\
\hline
\textbf{mix4}                      & \begin{tabular}{l}
prxy\_0, rsrch\_0 and mds\_0 
\end{tabular} &
\begin{tabular}{l}
All three workloads are write-intensive
\end{tabular}
& 56
\\
\hline
\textbf{mix5}                     & 
\begin{tabular}{l}
prxy\_0 and src2\_1
\end{tabular}
&
\begin{tabular}{l}
prxy\_0 is write-intensive and \\ 
src2\_1 is read-intensive
\end{tabular}
& 5
\\
\hline
\textbf{mix6}                     &   
\begin{tabular}{l}
prxy\_0, src2\_1 and usr\_0 
\end{tabular}
&
\begin{tabular}{l}
prxy\_0 is write-intensive, \\ 
src2\_1 is read-intensive and \\
usr\_0 \orevII{has 60\% writes and 40\% reads}
\end{tabular}
& 3
\\
\hline
\end{tabular}
}
\end{table}

\head{Metrics}
To compare \namePaper{} with prior systems, we report the following metrics in our experimental results (see \S\ref{sec:results}) \orevII{for each workload}\orevIV{:} 1) performance in terms of speedup in overall execution time \orevII{over \baseline{}}, 2) SSD throughput in IOPS (i.e., number of I/O operations per second), 3) tail \orevII{latency} in the 99th percentile of I/O requests, 4) SSD power/energy consumption, \orevII{ and 5) power and area overheads}.

\begin{figure*}[t]
\centering
 \includegraphics[width=0.95\textwidth]{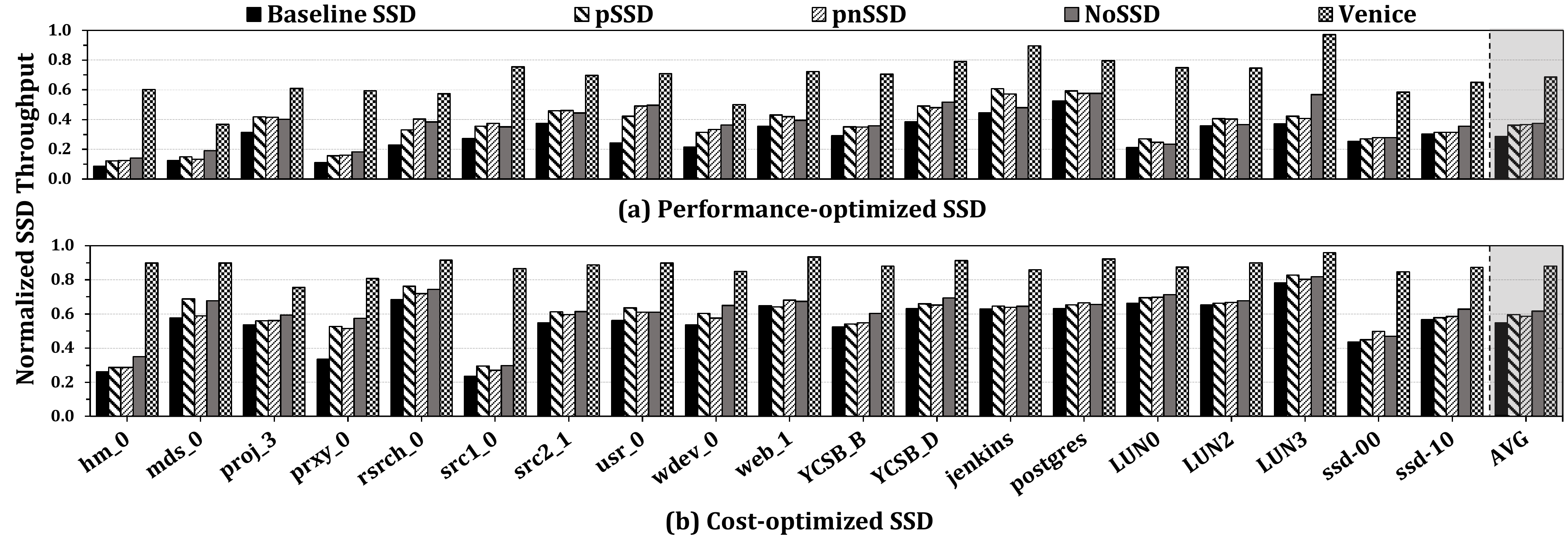}
 \vspace{-3mm}
 \caption{\revb{\orevII{SSD} Throughput of \baseline{}, \pssd{}, \pnssd{}, \nossd{} and \namePaper{} on performance-optimized \orevII{(top)} and cost-optimized \orevII{(bottom)} \orevIV{SSD} configurations. Throughput values (in IOPS) are normalized to the \ideal{}.}
 }
\label{fig:results_throughput}
\end{figure*}
\section{Evaluation \label{sec:results}}

\subsection{Performance Analysis\label{subsec:results_performance}}

\head{Execution Time}
Figures~\ref{fig:results_overall_speedup}(a) and \ref{fig:results_overall_speedup}(b) show the performance improvement of \pssd{}, \pnssd{}, \nossd{}, \namePaper{} and \ideal{} over \baseline{} in terms of speedup in overall execution time of each workload in a performance-optimized SSD and a cost-optimized SSD, respectively.

We make three key observations. First, \namePaper{} consistently outperforms \orevII{all} the prior approaches across all workloads in both SSD configurations. In the performance-optimized SSD configuration,  %
\namePaper{} outperforms \baseline{}/\pssd{}/\pnssd{}/\nossd{} by an average of  \orevII{2.65$\times$/2.10$\times$/2.00$\times$/1.92$\times$} across all workloads. In the cost-optimized SSD configuration,   %
\namePaper{} shows an average performance speedup of \orevII{1.67$\times$/1.52$\times$/1.55$\times$/1.47$\times$} over \baseline{}/\pssd{}/\pnssd{}/\nossd{}. 
Second, \namePaper{} results in higher performance improvements in the performance-optimized SSD configuration. This is because the performance-optimized SSD uses fast flash chips (with significantly lower read/write latencies), and thus the \vetII{I/O data} transfer time within the SSD dominates the I/O service time. \vetII{As a result, improving I/O data transfer performance in the performance-optimized SSD  provides a higher improvement in workload execution time.} Third, \namePaper{} performs within 45\% and 25\% of the \ideal{} in the performance-optimized and cost-optimized SSD configuration, respectively. 
We conclude that \namePaper{} significantly improves workload execution time by mitigating the \conf~problem in modern SSDs.

\head{\orevII{SSD} Throughput}
Figures~\ref{fig:results_throughput}(a) and \ref{fig:results_throughput}(b) show the SSD throughput (in IOPS) of \baseline{}, \pssd{}, \pnssd{}, \nossd{} and \namePaper{}  in a performance-optimized SSD and cost-optimized SSD, respectively. We normalize the SSD throughput results to the \ideal{}'s throughput.

We make two key observations. First, \namePaper{} improves \orevII{SSD} throughput over \baseline{}/\pssd{}/\pnssd{}/\nossd{} by 176\%/120\%/113\%/102\% in the performance-optimized SSD and 76\%/58\%/61\%/51\% in the cost-optimized SSD configuration. 
Second, \namePaper{}'s \orevII{SSD} throughput \orevII{is} within 30\% and 10\% of the \ideal{}'s throughput in the performance-optimized and cost-optimized SSD configuration, respectively. 
We conclude that \namePaper{} significantly improves SSD throughput by mitigating the \conf~problem.%

\begin{figure}[t]
\centering
\includegraphics[width=0.9\linewidth]{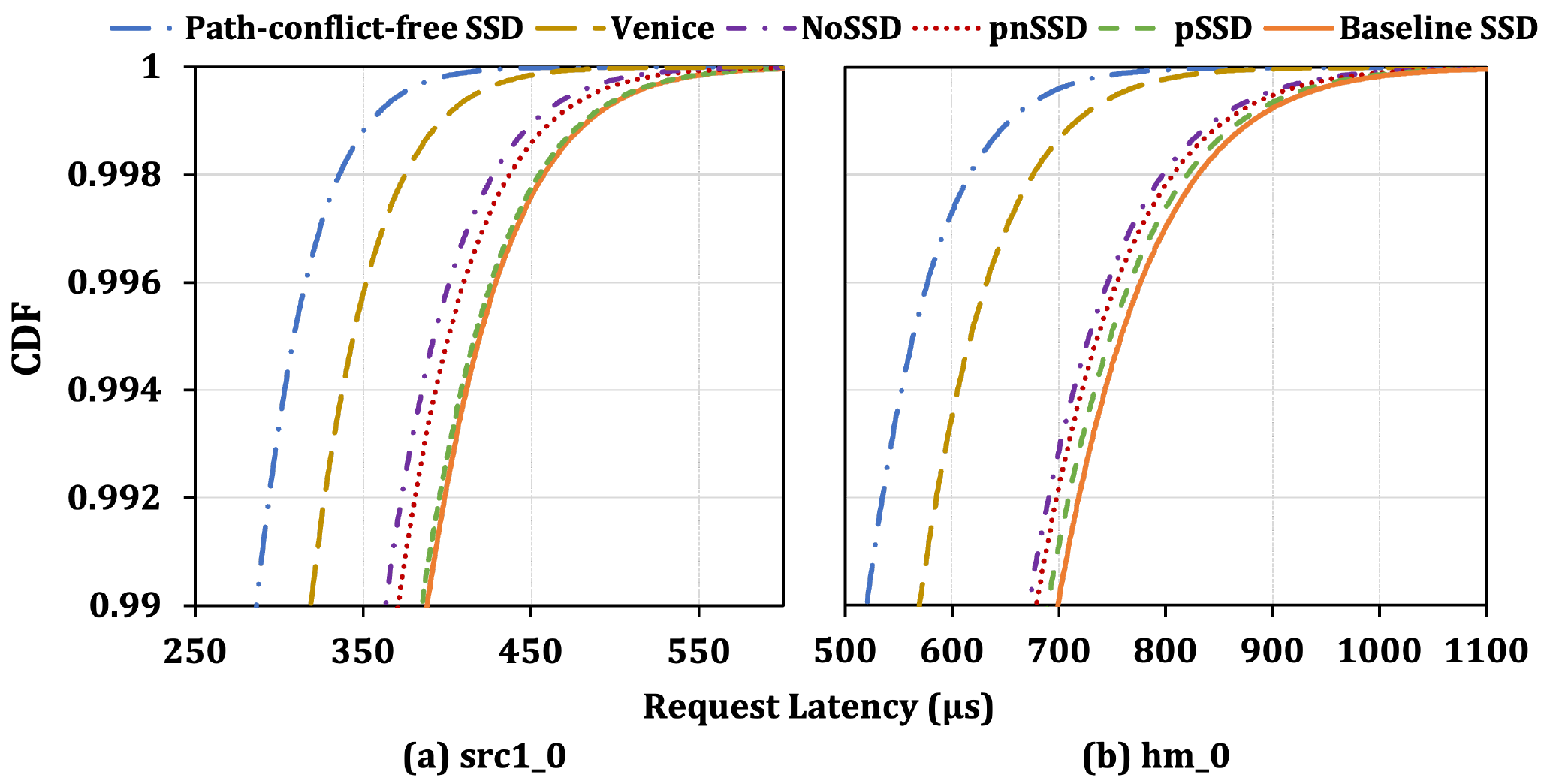}
\vspace{-4mm}
\caption{Comparison of tail latencies in the 99th percentile of I/O requests from two workloads, src1\_0 and hm\_0, on a performance-optimized SSD.}
\label{fig:result_tail_latency}
\end{figure}

\head{Tail Latency}
\Confs~can cause some I/O requests to experience significantly long access latencies. Figures~\ref{fig:result_tail_latency}(a) and~\ref{fig:result_tail_latency}(b) show the \orevII{99th percentile of I/O request latencies (i.e., tail latency)} in \orevIV{the} \orevIV{\ideal{}, \namePaper{}, \nossd{}, \pnssd{}, \pssd{} and \baseline{}} in the form of a cumulative density function (CDF) for two representative workloads \texttt{src1\_0} and \texttt{hm\_0},  respectively. We use only the performance-optimized SSD configuration for this experiment. We make the key observation that \namePaper{} significantly reduces the tail latency compared to prior systems. For \texttt{src1\_0}, \namePaper{} reduces the tail latency by 32\%/31\%/30\%/27\% over \baseline{}/\pssd{}/\pnssd{}/\nossd{}. For \texttt{hm\_0}, \namePaper{} reduces the tail latency by 22\%/21\%/18\%/17\% over \baseline{}/\pssd{}/\pnssd{}/\nossd{}.

\subsection{Mixed Workloads
\label{subsec:mixed_workloads}}
 
\begin{figure}[h]
\centering
\includegraphics[width=0.95\linewidth]{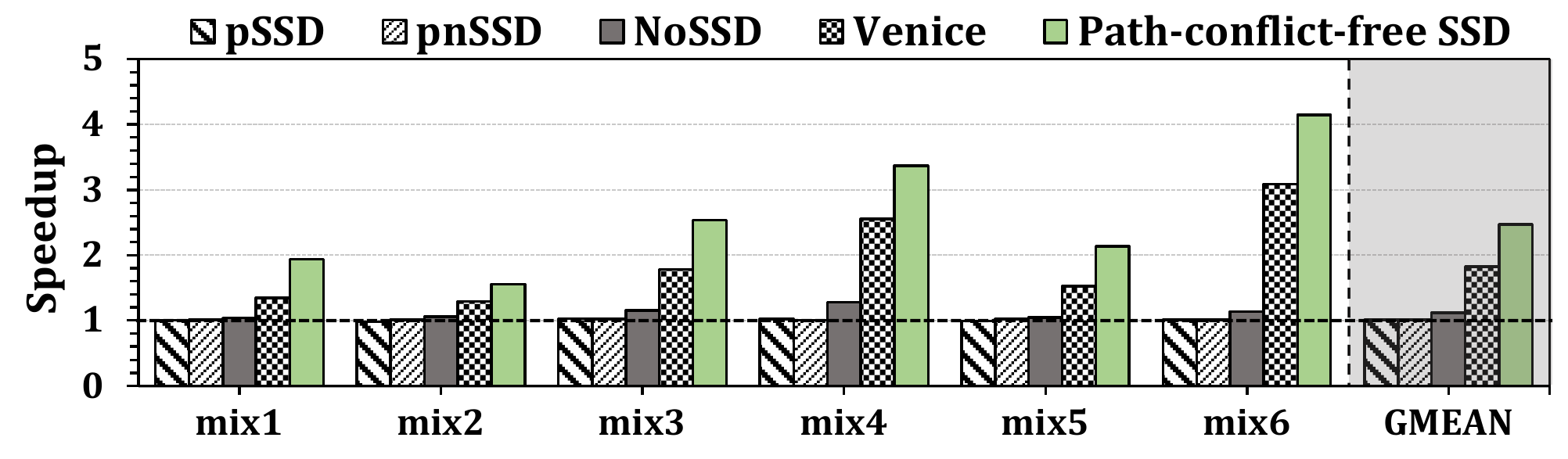}
\vspace{-2mm}
\caption{Performance comparison for mixed workloads on a performance-optimized SSD. \reva{Performance is measured in terms of speedup in overall execution time of each mixed workload over \orevII{the} \baseline{}}.}
\label{fig:mixed_workloads}
\end{figure}

To evaluate the effectiveness of \namePaper{} at improving SSD performance in real-world scenarios \orevII{where multiple workloads access the SSD}, we compare SSD performance using different systems under six mixed workloads.
Figure~\ref{fig:mixed_workloads} shows the speedup of \pssd{}, \pnssd{}, \nossd{}, \namePaper{} and \ideal{} over \baseline{} for six mixed workloads. We report \orevII{results} only for the performance-optimized SSD configuration.

\begin{figure*}[b]
\centering
 \includegraphics[width=0.98\textwidth]{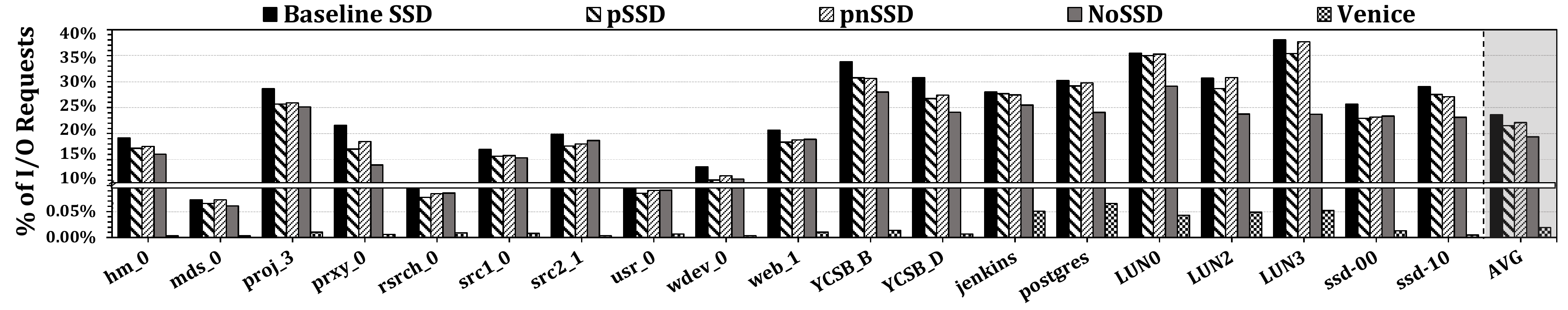}
 \vspace{-4mm}
 \caption{\dq{Percentage of I/O requests in \baseline{}, \pssd{}, \pnssd{}, \nossd{} and \namePaper{} that experience \confs{} in each workload on a performance-optimized SSD configuration. \orevIV{The y-axis} is shown in two parts in order to display the negligible number of I/O requests that suffer from \confs{} in \namePaper{}.} 
 }
\label{fig:results_blocked_requests_with_venice}
\end{figure*}

We make two key observations.
First, across all mixed workloads, \namePaper{} \orevII{improves} performance \orevII{over} prior \orevII{works}. \namePaper{} provides an average speedup of 1.83$\times$/1.81$\times$/1.80$\times$/1.63$\times$ over \baseline{}/\pssd{}/\pnssd{}/\nossd{}.
Second, \namePaper{}'s performance improvement is higher 
in \texttt{mix6}. In \texttt{mix6}, the average inter-request arrival time is 90\% lower than its constituent workloads, leading to increased \confs~in \baseline{}. \namePaper{} is able to schedule I/O requests on conflict-free paths using \orevII{its} non-minimal fully-adaptive routing algorithm. %
We conclude that \namePaper{} outperforms prior approaches on high-intensity mixed workloads by effectively mitigating path conflicts.

\subsection{Path \orevII{Conflict} Analysis
\label{subsec:results_blocked_requests}}
To show the effectiveness of \namePaper{} at mitigating the path conflict problem, we measure the percentage of I/O requests in each workload that experience path conflicts using different systems. 
Figure~\ref{fig:results_blocked_requests_with_venice} shows the \orevII{results} 
for the \baseline{}, \pssd{}, \pnssd{}, \nossd{} and \namePaper{} in the performance-optimized SSD configuration.

We make \orevII{the} key observation that \namePaper{} significantly mitigates the path conflict problem. Our experimental results show that \namePaper{} provides conflict-free paths (on the first try) for 99.98\% of I/O requests, on average, across all workloads, while \baseline{}/\pssd{}/\pnssd{}/\nossd{} provides conflict-free paths for 76.40\%/78.47\%/77.88\%/80.65\% of I/O requests. For a small number of I/O requests (i.e., 0.02\% of I/O requests, on average), \namePaper{} fails to find conflict-free paths on the first try, and thus, the corresponding I/O requests should wait for a longer amount of time until \namePaper{} successfully reserves conflict-free paths. \namePaper{}'s path reservation process can fail due to two major reasons.  
\orevII{%
First, if all flash controllers are busy processing ongoing I/O requests, \namePaper cannot start the path reservation process for a new I/O request until a flash controller becomes idle. 
Second, during the path reservation process, a \pkt{} \orevIV{(\orevV{see} \S \ref{subsec:mechanism_path_reserve})} cannot reserve a path if all the links leading to the destination flash chip are reserved.}
We conclude that \namePaper{} \orevII{effectively eliminates} the path conflict problem \orevII{via} path reservation and effective utilization of path diversity \orevII{in the SSD interconnection network}.

\subsection{Power and Energy Consumption}
\label{subsec:energy_analysis}
We study the impact of \namePaper{} and prior approaches on the SSD power and energy consumption.

\begin{figure*}[h]
\centering
 \includegraphics[width=0.95\textwidth]{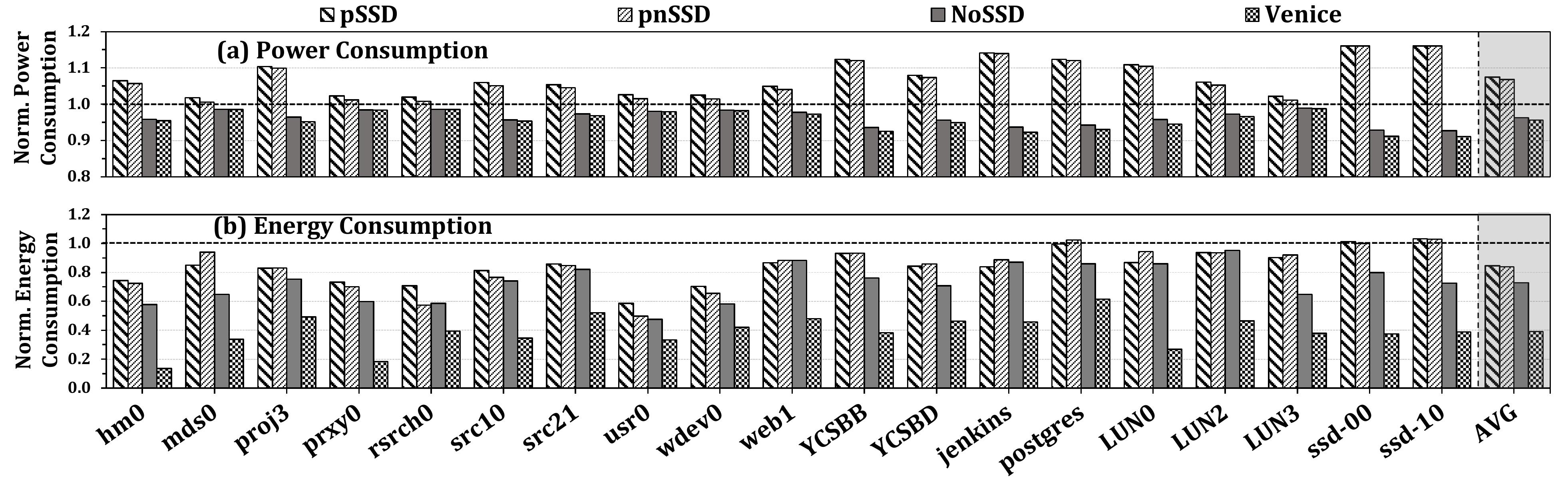}
 \vspace{-4mm}
 \caption{Power consumption \orevII{(top)} and \orevV{energy} \orevIV{consumption} \orevII{(bottom)} for \pssd{}, \pnssd{}, \nossd{} and \namePaper{} on the performance-optimized SSD configuration. \orevII{Power} and energy consumption values are normalized to \orevII{the} \baseline{}.
 }
\label{fig:power_energy_consumption}
\end{figure*}

\head{\orevII{Average} \orevIV{SSD} Power Consumption}
Figure~\ref{fig:power_energy_consumption}(a) shows the average  power consumption for \pssd{}, \pnssd{}, \nossd{} and \namePaper{} on a performance-optimized SSD configuration. \orevII{Average power} consumption values are normalized to \orevII{the average power consumption of the} \baseline{}. 
We make four key observations. 
First, \namePaper{} reduces SSD average power consumption by 4\% compared to the \baseline{}. This is mainly because a link in the interconnection network consumes significantly lower power than the shared channel (see \S\ref{subsec:overhead}). 
Second, \namePaper{} consumes slightly less power (around 1\%)  than \nossd{} due to \namePaper{}'s simple router design. 
Third, the impact of \namePaper{} and prior systems on SSD power consumption is \orevII{small} since the SSD power consumption is dominated by flash operations (i.e., read, program, and erase). The number of flash operations remains the same in \namePaper{} and all prior systems. 

\head{\orevIV{SSD} Energy Consumption}
To calculate the energy consumption of \namePaper{} and prior approaches, we multiply the average power consumption by the overall execution time of each workload. 
Figure~\ref{fig:power_energy_consumption}(b) shows the energy consumption for \pssd{}, \pnssd{}, \nossd{} and \namePaper{} on a performance-optimized SSD configuration. \orevII{Energy} consumption values are normalized to \orevII{the} \baseline{}. 

We make \orevII{the} key observation that \namePaper{} has significantly lower energy consumption than prior approaches across all workloads. \namePaper{} reduces energy consumption by an average of 61\%/54\%/53\%/46\% compared to \baseline{}/\pssd{}/\pnssd{}/\nossd{}. We conclude that \namePaper{}'s lower \orevII{average} power consumption and \orevII{lower} execution time \orevIV{together} result in \orevII{largely} \orevIV{lower} energy \orevII{consumption} compared to prior systems.

\subsection{Sensitivity to \orevIV{Interconnection Network Configurations}\label{subsec:results_sensitivity}}
\begin{figure}[h]
\centering
\includegraphics[width=0.98\linewidth]{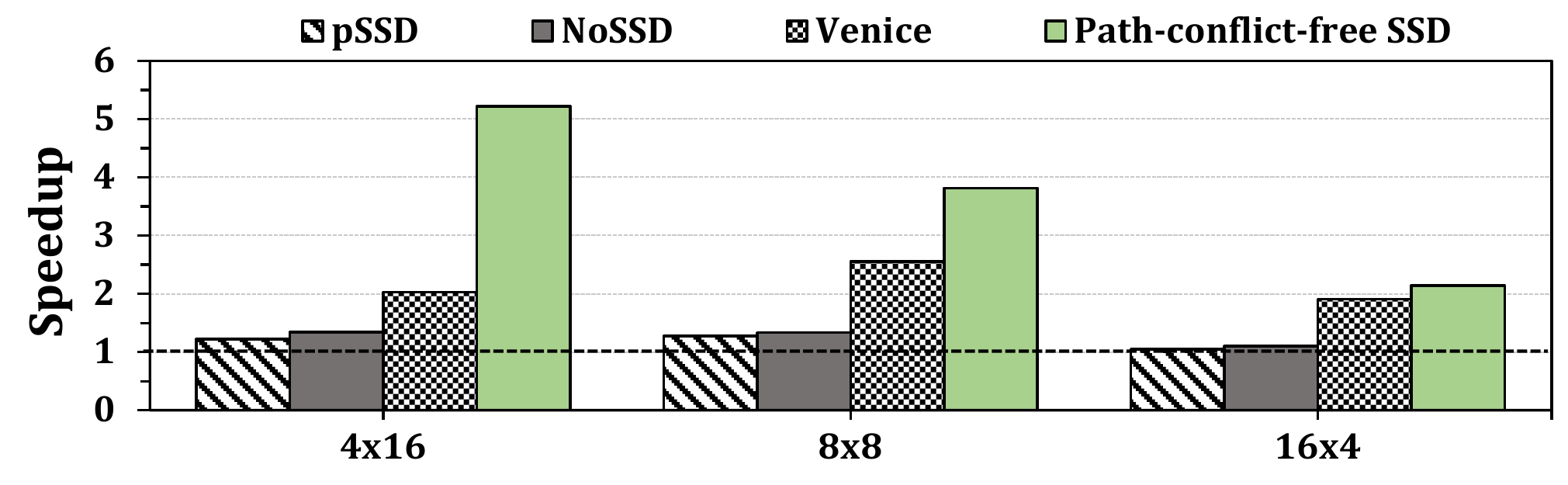}
\vspace{-3mm}
\caption{Performance speedup of \pssd{}, \nossd{}, \namePaper{} and \ideal{} in the performance-optimized SSD configuration. The X-axis shows three \orevII{systems} with 4, 8 and 16 flash controllers respectively. For example, 4$\times$16 denotes 4 flash controllers and 16 flash chips connected in each row of the flash array. The speedup in overall execution time over \baseline{} is averaged across all workloads.}
\label{fig:sensitivity_channel_chips}
\end{figure}

We study the \orevII{effect} of 
\vetII{the interconnection network configuration} %
on \orevII{\namePaper{}'s} performance improvement. To this end, we compare the performance of \orevII{\namePaper{} and prior works} using three \orevII{systems that use} 4, 8, and 16 flash controllers in the performance-optimized SSD configuration. We keep the total number of flash chips in the SSD constant across the three \orevII{systems}. Figure~\ref{fig:sensitivity_channel_chips} shows the \orevIV{average} speedup of \pssd{}, \nossd{}, \namePaper{}, and \ideal{} over the \baseline{}, across all workloads. The X-axis shows three \orevII{systems}, 4$\times$16, 8$\times$8, and 16$\times$4. 4$\times$16, for example, denotes four flash controllers with 16 flash chips in each row of the flash array.\footnote{Note that we omit \pnssd{} from this study because \pnssd{} requires an N$\times$N flash array configuration where N is the number of flash controllers as well as the number of flash chips in each row in the flash array. Hence, 4$\times$16 and 16$\times$4 configurations are \orevII{\emph{not}} supported by \pnssd{}.} 

We make two key observations from our sensitivity analysis. First, \namePaper{} provides significant speedup over prior approaches across all three \orevII{systems with different numbers of flash controllers}. 
\namePaper{} outperforms \baseline{}/\pssd{}/\nossd{} by 1) 2$\times$/1.7$\times$/1.5$\times$ in  4$\times$16, 2) 2.6$\times$/2$\times$/1.9$\times$ in 8$\times$8, and 3) 1.9$\times$/1.8$\times$/1.7$\times$ in 16$\times$4. 
Second, \namePaper{}'s performance improvement is higher for the 8$\times$8 flash array configuration compared to both 4$\times$16 and 16$\times$4 configurations. 
\vetII{In the system with 4 flash controllers, \namePaper{} can reserve conflict-free paths for up to four ongoing I/O requests. \namePaper{} can reserve conflict-free paths for up to eight I/O requests in the system with 8 flash controllers. As a result, \namePaper{} has a lower ability to eliminate the \conf~problem in the system with 4 flash controllers, which results in lower performance improvements \orevIV{for \namePaper{}}. %
On the other hand, the system with 16 flash controllers has \orevVI{a} lower \orevV{number of} \confs~compared to \orevIV{systems with 4 and 8 flash controllers}, and thus, \namePaper{} provides lower performance \orevV{improvement} \orevIV{compared to those in the other two systems}. \orevIV{We} conclude that \namePaper{} is effective for different SSD interconnection network configurations. 
}

\dq{
\subsection{\orevII{Power and Area} Overhead Analysis}
\label{subsec:overhead}

\noindent\textbf{Power.} As discussed in \S\ref{subsec:energy_analysis}, \namePaper{} reduces average power and energy consumption. This section studies the power consumption of 
\nika{\namePaper{}'s interconnection network router and links}.
\nika{We analyze the power consumption of the router by implementing its}
hardware description language (HDL) model and synthesizing \nika{it} for \orevII{the} UMC 65nm technology node\orevII{~\cite{umc}}. We observe that each router consumes 0.241 mW. 
\nika{We \orevII{measure the power consumption of} each network link using} ORION 3.0~\cite{kahng2015orion3} power model tool and observe that each link consumes about 1.08 mW for a 4KB NAND flash page transfer, which is 90\% less power consumption than that of a shared channel bus. 
Each network link consumes significantly lower power than a shared channel bus due to its lower capacitance load. Link capacitance is lower than bus capacitance since (1) it is shorter and thinner than a shared bus and (2) it has only two drivers  \nika{compared to several (e.g., 8 in an $8\times$8 flash array configuration) drivers in a shared bus}. 
\orevII{Table \ref{tab:overhead} (3rd column) summarizes  the power consumption of \namePaper{}'s components.} \vetII{We have already observed in \S\ref{subsec:energy_analysis} that \namePaper{} reduces the average power consumption \orevIV{by 4\% over} \baseline{}}.

\begin{table}[h]
 \caption{Power \orevIV{and \orevV{area overheads}} of \namePaper{}}
 \vspace{-3mm}
    \label{tab:overhead}
\centering
\normalsize
 \setstretch{1}
   \renewcommand{\arraystretch}{1} %
\setlength{\tabcolsep}{2pt}
  \resizebox{\linewidth}{!}
{
\begin{tabular}{|c||c|c|c|}
\hline
\textbf{Component} &  
\begin{tabular}{c}    
\textbf{\# of Instances} 
\end{tabular} &
\begin{tabular}{c} 
\textbf{Avg. Power [$mW$] } \\ 
\textbf{for 4KB page transfer}
\end{tabular} &  
\begin{tabular}{c}    
\textbf{Area} 
\end{tabular} \\
\hline
\hline
\textbf{Router} &
\begin{tabular}{c}    
1 per flash node 
\end{tabular} &
0.241 & 
\begin{tabular}{c}    
8\% of flash chip area
\end{tabular}
\\
\hline
\textbf{Link} &
\begin{tabular}{c}    
Up to 4 per flash node 
\end{tabular}  &
1.08 &
\begin{tabular}{c}    
0.04$\times$ flash channel area
\end{tabular}
\\
\hline
\end{tabular}
}
\end{table}

\noindent\textbf{Area.} \namePaper{} does \emph{not} impose area overhead on NAND flash chip design as \nika{it} does \orevII{\emph{not}} integrate the router inside the NAND flash chip \orevII{(i.e., no pins are added to the commodity flash chips)}.
Router chips and the links connecting them can impose area overhead on SSD printed circuit board (PCB) design. \nika{To estimate this overhead, we model the area overhead of the interconnection network's routers and links.} 

\nika{We} estimate the area overhead of \namePaper{}'s routers \nika{using} the HDL model of the router. 
Each router in \namePaper{} has an area of 614 $\mu m^{2}$. However, each router occupies a higher area on \orevII{the} PCB due to its I/O pad overheads. Each router has 40 I/O pins. Considering I/O pad sizes (about 0.2 mm) and the safety distance between two I/O pads (about 0.2 mm), we expect that each router occupies about 8 $mm^{2}$, which is 8\% of \orevII{a} typical 100 $mm^{2}$ NAND flash chip~\cite{kang201913}. 

\nika{We estimate the area overhead of \namePaper{}'s links using ORION 3.0~\cite{kahng2015orion3}.} 
Assuming an 8$\times$8 2D mesh topology connecting 64 NAND flash chips, \namePaper{} requires 112 network links \nika{(instead of the eight \orevII{shared channels} required in the baseline).} 
Note that we do not count injection/ejection links as they are the same as flash chips' connectors to the shared channel bus. 
\nika{Our experimental results show that each link's area is roughly 0.04$\times$ of the shared channel area. As a result, in total, \namePaper{}'s interconnect links \vetII{occupy} 44\% lower area compared to the baseline multi-channel shared bus architecture.}\footnote{\orevV{ We measure the total area overhead of \namePaper{}'s interconnect links \orevVI{compared to a baseline multi-channel shared bus architecture} using this equation:\\
\orevVI{1 - (total area of interconnection network links) / (total area of channels in multi-channel shared bus architecture), i.e.,} 1 - (\#Links $\times$ $Link_{area}$) / (\#Channels~ $\times$ $Channel_{area}$) = \,1 - (112~$\times$ 0.04) / (8 $\times$ 1) = 0.44}.}
A network link occupies significantly \orevII{smaller} space than a bus for two main reasons. First, links are shorter than buses (e.g., by 8$\times$ in our case); thus, link wires can be 8$\times$ thinner to ensure the same impedance as the bus. Second, as the link wires are thinner, they require lower pitch sizes, reducing the \nika{overall area required by the} links. 
\orevII{We summarize the area overhead of  \namePaper{}'s components in Table \ref{tab:overhead} (4th column).} \orevII{We conclude that \namePaper{}'s benefits come at relatively low area overhead.}}

\section{Related Work\label{sec:relatedwork}}
To our knowledge, \namePaper{} is the first work that fundamentally addresses the \conf~problem in SSDs at low cost. %
We have quantitatively compared \namePaper{} extensively to three \orevIV{major} prior works, \pssd{}~\cite{kim2022networked}, \pnssd{}~\cite{kim2022networked} and \nossd{}~\cite{tavakkol2012network, tavakkol2014design} in \S\ref{sec:results}. In this section, we briefly review related work in two domains: 1) improving \orevII{flash array parallelism}, and 2) \orevII{exploiting flash array parallelism}.

\head{Improving \orevII{Flash Array Parallelism}} 
\orevIV{Prior} works propose to employ \nika{an} interconnection network inside the SSD (e.g.,~\cite{kim2022networked,tavakkol2012network,tavakkol2014design,schuetz2007hyperlink,gillingham2013800,gillingham2011256gb,kim2021decoupled}).   
HyperLink NAND flash architecture (HLNAND)~\cite{schuetz2007hyperlink, gillingham2013800, gillingham2011256gb} %
connects the flash chips using a ring-topology interconnection network. Decoupled SSD~\cite{kim2021decoupled} proposes an on-chip router within each flash controller to create a network of flash controllers in the SSD. %
Unfortunately, both HLNAND and Decoupled SSD do \emph{not} provide rich path diversity between the flash controllers and flash chips, and thus, cannot effectively mitigate the \conf~problem.

\head{\orevII{Exploiting Flash Array} Parallelism} 
\orevIV{Other prior} works (e.g.,~\cite{park2010exploiting, ruan2012improving, jung2012physically,mao2017improving, gao2019parallel, jung2012evaluation, gao2014exploiting, gao2017exploiting,tavakkol2018flin,wang2013novel}) attempt to \orevIV{exploit the} internal parallelism \orevIV{in an SSD to improve the SSD performance}.
\orevIV{These works mainly focus on} I/O scheduling. 
Jung et al.~\cite{jung2012physically} propose Physically Addressed Queueing (PAQ), an I/O scheduler implemented in a layer between the FTL and the flash array. PAQ selects groups of operations that can be simultaneously executed without \orevII{contention for a shared resource (e.g., flash channel)}. 
Gao et al.~\cite{gao2014exploiting, gao2017exploiting} propose Parallel Issue Queuing (PIQ), a host I/O scheduler that batches I/O requests that use different flash channels to be scheduled simultaneously to exploit SSD-level parallelism. 
\orevVI{FLIN~\cite{tavakkol2018flin} provides} both high-performance and fair I/O scheduling \orevVI{in modern SSDs}. We believe \namePaper{} is orthogonal to these works and \orevVII{I/O scheduling for the \namePaper{} architecture} is an interesting research direction.

\orevIV{Several} prior works~\cite{hu2011performance, hu2012exploring, tavakkol2016performance, jung2012evaluation, chen2011essential, xie2018exploiting, zhang2019spa} propose \orevII{techniques to exploit flash array parallelism} by focusing on physical page allocation in the FTL. Unfortunately\nika{,} these works fail to effectively lay out data such that SSD does not experience the \conf~problem. This is due to 1) random data access pattern\nika{s} in SSDs, 2) I/O interference from multiple concurrent applications, and 3) dynamic changes in device conditions.

\section{Discussion\label{sec:discussion}}
We propose \namePaper{} to \nika{improve SSD performance by } mitigating the path conflict problem in SSDs. \namePaper{} can be extended to improve SSD and system performance in other ways. We discuss \nika{other} use cases of \namePaper{} in this section.

\head{Applicability of \namePaper{} to Near-Data Processing (NDP)}
NDP is a computing paradigm that moves the computation closer to where the data resides \orevV{(e.g.,~\cite{seshadri-ieeecal-2015, seshadri-micro-2017, hajinazar-asplos-2021, seshadri-osdi-2014, gao-micro-2021, mutlu-emergingcomputing-2021, park2022flash,aga-hpca-2017,li-dac-2016,mansouri-asplos-2022,gu-isca-2016,mailthody-micro-2019,mutlu2019processing,pei-tos-2019,truong-micro-2021,acharya-asplos-1998,koo-micro-2017,kang-micro-2021,jun-isca-2018,ghose2019processing,riedel2001active,riedel1998active,kang2013enabling,keeton1998case,do2013query,kim2016storage,torabzadehkashi2019catalina,lee2020smartssd,ajdari2019cidr})}. 
\nika{\namePaper{}'s improved parallelism can facilitate NDP inside the SSD by efficiently co-locating different operands required by the NDP operations.}
Prior proposals~\cite{gao-micro-2021, park2022flash} that perform in-flash bulk bitwise operations have data \orevII{location} constraints where the operands must be moved to a single flash chip before the computation is performed. Path conflicts can impact this data movement, which can significantly reduce the performance benefits of in-flash processing. 
\namePaper{} \nika{can leverage its improved flash-array parallelism to efficiently}
gather operand data from different flash chips \nika{to the target flash chip that performs the NDP computation.}

\head{Improving Garbage Collection}
The garbage collection (GC) process\orevII{~\cite{yang2014garbage, cai2017error, tavakkol2018mqsim, agrawal2008design, shahidi2016exploring, lee2013preemptible,jung2012taking, choi2018parallelizing,wu2016gcar,cai-errors-2018,cai2017vulnerabilities}} in NAND flash-based systems is critical to reduce fragmentation and maintain free blocks for write operations. During GC, the SSD controller reads a large number of valid pages from victim blocks. These pages are written to new blocks in the same flash chip or a different flash chip. This data movement can interfere with I/O requests\orevIV{~\cite{kim2019alleviating,wu2016gcar,tavakkol2018flin,tsai2019learning,tavakkol2018mqsim,kim2022networked,kim2021decoupled,chen2016internal}} and cause path conflicts.
\nika{\namePaper{} can leverage its improved path diversity to efficiently} schedule both host I/O requests and GC-related requests in parallel.

\section{Conclusion\label{sec:conclusion}}
We \orevII{propose} \namePaper{}, a new mechanism that introduces a low-cost interconnection network of flash chips \nika{and utilizes the path diversity
efficiently} to fundamentally address the path conflict problem in SSDs.  
\namePaper{} mitigates path conflicts and improves \vetII{SSD} parallelism using three key techniques\nika{:} (1) a simple router chip placed next to each flash chip without modifying the flash chip itself, (2) a path reservation technique to reserve a path for each I/O request from the SSD controller to the target flash chip, and (3) a non-minimal fully-adaptive routing algorithm to effectively utilize the path diversity in the interconnection network. 
Our evaluation shows that \namePaper{} significantly improves performance over state-of-the-art prior approaches on both \orevIV{performance-optimized} and cost-optimized SSD configurations \nika{for} a wide range of real-world data-intensive workloads\orevII{, by effectively eliminating \confs{}}. 
As the demand for performance and scalability of SSDs increases, we hope that \namePaper{} inspires future work in several directions to mitigate path conflicts and improve parallelism within the SSD.

\begin{acks}
We thank the anonymous reviewers of ISCA 2023 for their feedback and comments.
We thank the SAFARI Research Group members for valuable feedback and the stimulating intellectual environment they provide. 
We acknowledge the generous gifts of our industrial partners, especially Google, Huawei, Intel, Microsoft and VMware. This research was partially supported by the Semiconductor Research Corporation, the Swiss National Science Foundation, and the ETH Future Computing Laboratory.
\end{acks}

\newpage
{
\balance
\bibliographystyle{IEEEtran}

}

\end{document}